\documentclass[
aps,
prd,
reprint,
superscriptaddress,
nofootinbib,
floatfix
]{revtex4-2}

\usepackage{amsmath,amssymb,amsfonts}
\usepackage{graphicx}
\usepackage{hyperref}

\usepackage{float}
\usepackage{mathtools}          % \coloneqq, aligned, etc.
\usepackage{geometry}
\usepackage{subcaption}
\usepackage[english]{babel}
\usepackage{xcolor} 
\usepackage[normalem]{ulem}
\usepackage{tikz}
\usetikzlibrary{positioning,arrows.meta}

\begin{document}

\title{Towards solving General Relativity with Physics-Informed Neural Networks}

\author{Juan A. Carretero}
\email{juan-antonio.carretero@uib.cat}
\affiliation{Departament de F\'isica, Universitat de les Illes Balears, Palma de Mallorca, E-07122, Spain}
\affiliation{Institute of Applied Computing \& Community Code (IAC3),
 Parc BIT - Complex Balear de Recerca i Desenvolupament Tecnològic, Palma de Mallorca, E-07122, Spain}

\author{Jorge F. Urbán}
\email{jorgefrancisco.urban@ua.es}
\affiliation{Departament de Física, Universitat d'Alacant, Ap. Correus 99, E-03080 Alacant, Spain}

\author{Fernando Abalos}
\affiliation{Departament de F\'isica, Universitat de les Illes Balears, Palma de Mallorca, E-07122, Spain}
\affiliation{Institute of Applied Computing \& Community Code (IAC3),
 Parc BIT - Complex Balear de Recerca i Desenvolupament Tecnològic, Palma de Mallorca, E-07122, Spain}

\author{Joan Massó}
%\email{joan.masso@uib.es}
\affiliation{Departament de F\'isica, Universitat de les Illes Balears, Palma de Mallorca, E-07122, Spain}
\affiliation{Institute of Applied Computing \& Community Code (IAC3),
 Parc BIT - Complex Balear de Recerca i Desenvolupament Tecnològic, Palma de Mallorca, E-07122, Spain}

\author{Oscar A. Reula}
\affiliation{Facultad de Matem\'atica, Astronom\'ia, F\'isica y Computaci\'on, Universidad Nacional de C\'ordoba, Argentina} \affiliation{ Instituto de F\'isica Enrique Gaviola, CONICET, C\'ordoba, Argentina}

\author{Jose A. Pons}
%\email{jose.pons@ua.es}
\affiliation{Departament de Física, Universitat d'Alacant, Ap. Correus 99, E-03080 Alacant, Spain}

\author{Carlos Palenzuela}
\affiliation{Departament de F\'isica, Universitat de les Illes Balears, Palma de Mallorca, E-07122, Spain}
\affiliation{Institute of Applied Computing \& Community Code (IAC3),
 Parc BIT - Complex Balear de Recerca i Desenvolupament Tecnològic, Palma de Mallorca, E-07122, Spain}

\begin{abstract}
Physics-Informed Neural Networks (PINNs) are a machine-learning framework for approximating solutions to systems of partial differential equations by constraining neural networks to satisfy the underlying physical laws. 
The resulting continuous representation does not require a predefined computational mesh and can be evaluated at arbitrary points within the training domain. 
In this work, we investigate the application of PINNs to Einstein's equations of General Relativity. We discuss the mathematical formulation, network architectures, loss functions, and training strategies used to obtain accurate spacetime solutions, and demonstrate the approach through a series of standard benchmarks from Numerical Relativity. We then extend the method to a more astrophysical setting by considering the dynamical evolution of an isolated solitonic boson star. 
Our simulations show that 
a PINN operating directly in three-dimensional Cartesian coordinates can reproduce the equilibrium structure of a highly compact boson star and recover its characteristic radial oscillation frequencies.
These results establish PINNs as a viable complementary framework for solving coupled matter-Einstein equations and exploring alternative formulations in numerical relativity.
\end{abstract}

\maketitle

%%%%%%%%%%%%%%%%%%%%%%%%%%%%%%%%%%%%%
%%%%%%%%%%%%%%%%%%%%%%%%%%%%%%%%%%%%%
\section{Introduction}
%%%%%%%%%%%%%%%%%%%%%%%%%%%%%%%%%%%%%
%%%%%%%%%%%%%%%%%%%%%%%%%%%%%%%%%%%%%

In General Relativity, the geometry of spacetime is governed by Einstein's equations, a nonlinear system of ten coupled partial differential equations that describes phenomena such as black holes and gravitational waves~\cite{carroll,wald}. Our understanding of gravity has been greatly advanced by exact analytical solutions, perturbative approaches, and post-Newtonian approximations. 
However, these approaches are generally insufficient to describe the full evolution of strongly nonlinear systems, such as merging compact-object binaries. These systems involve highly dynamical and strongly curved spacetimes, as well as event horizons and relativistic matter, making numerical approaches essential.
Numerical relativity provides the framework to solve Einstein's equations in these regimes and, in particular,  to simulate compact-object binaries throughout their inspiral, merger, and ringdown phases~\cite{pretorius2005,campanelli2006}, which are of central importance in the era of gravitational-wave astronomy.

In numerical relativity, Einstein's equations are typically reformulated as an initial-value problem using a $(3+1)$ decomposition of spacetime (see, e.g., ~\cite{librocarlos}). This introduces additional challenges beyond the intrinsic nonlinearity of the field equations. In particular, the resulting evolution system is subject to constraints that must be satisfied on each spatial hypersurface, while the general covariance of General Relativity introduces gauge freedom associated with the choice of spacetime coordinates, which must be controlled through suitable gauge conditions. A successful numerical scheme must therefore evolve the physical degrees of freedom accurately while preserving the constraints and maintaining a consistent gauge throughout the evolution. To address these challenges, finite-difference, finite-volume, and spectral methods have become standard approaches for obtaining accurate numerical solutions to the equations governing relativistic systems (see, e.g., \cite{alcubierre,font2008,grandclement2009}).

In this work, we investigate Physics-Informed Neural Networks (PINNs) as a complementary approach to solving Einstein's equations coupled to matter. Our aim is not to replace established numerical-relativity codes, 
but to develop and systematically validate a neural representation of spacetime and matter fields
across qualitatively different physical regimes. 
For the problems considered here, training PINNs remains more computationally expensive than performing the corresponding finite-difference evolutions.
They nevertheless offer other potentially useful features, including flexibility and, in particular, the ability to represent continuous families of solutions rather than a single solution for a given set of physical parameters. Several extensions of the standard PINN framework have been proposed to enhance accuracy and reduce training costs; see, e.g., Ref.~\cite{toscano2025pinns} for a recent review.

PINNs represent the unknown fields using differentiable neural networks, which act as nonlinear function approximators with trainable internal basis functions.
The network parameters are optimized by minimizing a loss function constructed from the residuals of the governing equations, together with the prescribed initial, boundary, and gauge conditions ~\cite{original,resurgir, karniadakis2021physics, urban2025unveiling}.
For the second-order Einstein--Klein--Gordon (EKG) system considered here, the network takes the four spacetime coordinates as inputs and predicts twelve outputs: the ten independent components of the metric and the real and imaginary parts of the complex scalar field. The field equations are formulated directly as ten Einstein equations, four gauge conditions, and two Klein--Gordon equations for the scalar-field components. All derivatives are computed using automatic differentiation \cite{ad}. 
This formulation allows the covariant field equations to be enforced directly, without recasting the problem as a \(3+1\) evolution system with separately evolved auxiliary variables.

The application of PINNs to General Relativity and relativistic astrophysics is still in its infancy, with only a limited number of studies reported to date. Early work focused on simplified relativistic systems, demonstrating that PINNs can accurately reproduce known analytical solutions, such as the Schwarzschild spacetime \cite{Li2023}. 
Subsequent studies extended this approach to black hole perturbation theory by solving the Teukolsky equation on a Schwarzschild background \cite{Luna2023,Patel2024}. More recently, PINNs have also been applied to post-Newtonian binary dynamics \cite{Barbagallo2026}, where the orbital evolution is described by a system of ordinary differential equations. The authors model both conservative and dissipative effects, including radiation reaction, and construct parametric models capable of representing families of binary orbits. They have also been applied to a dynamical, fully non-linear relativistic problem: the gravitational collapse of a massless scalar field \cite{collapse}. In that work, the Einstein--Klein--Gordon equations were formulated under the assumptions of spherical symmetry and polar--areal coordinates, reducing the system to an effectively (1+1)-dimensional problem involving only the temporal and radial coordinates.

These preliminary studies have demonstrated that PINNs are capable of producing accurate solutions for spherically symmetric spacetimes using numerical techniques that differ substantially from conventional approaches. The framework presented here extends these previous works in several important respects. First, it employs a quasi-Newton optimization method to minimize the loss function \cite{urban2025unveiling}, together with a multi-window temporal decomposition strategy that enables accurate solutions over long evolution times \cite{multi,penwarden2023unified}. More importantly, it is formulated in the full four-dimensional setting and does not impose any symmetry reduction on either the field equations or the neural representation. Although several of the benchmark problems considered in this paper possess spatial symmetries, these are properties of the particular solutions rather than assumptions built into the formulation. 
Consequently, extending the implementation to genuinely multidimensional configurations does not require the reformulation of the underlying field equations.

We assess the accuracy and robustness of the proposed framework using a sequence of increasingly demanding benchmark tests.
We first consider the standard \textit{Apples with Apples} tests for numerical relativity \cite{mexico}, which provide 
analytical reference solutions probing
different aspects of relativistic dynamics, including wave propagation, gauge dynamics, and nonlinear, multiscale evolution. In particular, the gauge-wave test is formulated in a parametric form, allowing us to assess the ability of the PINN to learn and reproduce a continuous family of solutions. Together, these benchmarks provide a systematic assessment of the framework against known analytical solutions and test its ability to accurately capture their evolution over long timescales.

We then consider solitonic boson stars within the EKG system, one of the simplest frameworks for studying the nonlinear coupling between spacetime geometry and matter. Despite its simple matter content, consisting of a complex scalar field, the EKG system admits regular, self-gravitating compact objects known as boson stars~\cite{revBS}. For suitable self-interaction potentials, solitonic configurations can reach compactness comparable to, or even exceeding, that of neutron stars~\cite{solbs}, while remaining regular and horizonless. They therefore provide a useful testbed for numerical methods in non-vacuum, strong-field regimes. 

We evolve a solitonic boson star using a neural representation that depends on the full set of Cartesian spacetime coordinates, simultaneously predicting the metric and complex scalar field
that sources it through its energy--momentum tensor. We assess the evolution by comparing the resulting configuration with the expected stationary solution and an independent finite-difference evolution, while monitoring the conservation of the Noether charge and the Hamiltonian constraint residual. We further examine the oscillation spectrum of the star to assess whether the PINN accurately captures its characteristic modes. This test probes the ability of the neural representation to capture the coupled dynamics of matter and geometry while maintaining a regular, self-gravitating configuration over an extended evolution.

The remainder of this paper is organized as follows. Section~\ref{sec:formulation} presents the Einstein--Klein--Gordon system of equations, whose residuals are enforced during the training of the neural network. Section~\ref{sec:pinns} describes the PINN representation and the corresponding training strategy. Section~\ref{sec:tests} presents the vacuum benchmark tests, while Section~\ref{sec:boson_star} is devoted to the evolution of a solitonic boson star. Finally, Section~\ref{sec:conclusions} summarizes our main results and discusses the capabilities, limitations, and future prospects of the proposed approach.

%%%%%%%%%%%%%%%%%%%%%%%%%%%%%%%%%%%
%%%%%%%%%%%%%%%%%%%%%%%%%%%%%%%%%%%
\section{The Einstein-Klein-Gordon  equations }
\label{sec:formulation}
%%%%%%%%%%%%%%%%%%%%%%%%%%%%%%%%%%%
%%%%%%%%%%%%%%%%%%%%%%%%%%%%%%%%%%%

In this section, we present the fully covariant formulation of the EKG system underlying our approach and specify the 16 partial differential equations whose residuals are enforced during PINN training. Throughout the paper, we use geometrized units with $G=c=1$. Spacetime indices are denoted by $a,b,c,..=0,1,2,3$, while spatial indices are denoted by $i,j,k,..=1,2,3$.

%%%%%%%%%%%%%%%%%%%%%%%%%%%%%%%%%%
\subsection{Einstein equations}
%%%%%%%%%%%%%%%%%%%%%%%%%%%%%%%%%%

The spacetime metric $g_{ab}$ is determined by the Einstein equations, which, in geometrized units, take the form
\begin{eqnarray}
    R_{ab} &=& 8\pi ( T_{ab}   -
    \frac{1}{2}T g_{ab})
    \label{eq:einstein} 
\end{eqnarray}
where $T=g^{ab} T_{ab}$ is the trace of the stress-energy tensor and the Ricci tensor $R_{ab}$ is given by
\begin{eqnarray}
    R_{ab}  &=&
    -\frac{1}{2} g^{cd} \partial_c\partial_d g_{ab}
    +  \frac{1}{2}\partial_{a}\Gamma_{b}
    +  \frac{1}{2}\partial_{b}\Gamma_{a}
    \\
    &-&   \Gamma^c\Gamma_{cab}
    +  g^{cd}g^{ef}
    \left(
        \partial_e g_{ca}\,
        \partial_f g_{db}  -
        \Gamma_{ace}\Gamma_{bdf}
    \right) .
    \nonumber
\label{eq:ricci_decomposition}
\end{eqnarray}
Here $\Gamma^a{}_{bc}$ are the Christoffel symbols, and
\begin{equation}
    \Gamma^a
    \equiv
    g^{bc}\Gamma^a{}_{bc},
    \qquad
    \Gamma_a=g_{ab}\Gamma^b~.
\end{equation}
denote their contractions, which are introduced to simplify the expression for the Ricci tensor.

The general covariance of Einstein's equations implies that a given physical spacetime admits infinitely many mathematically distinct representations, related by coordinate transformations. Consequently, Einstein's equations alone do not uniquely determine the evolution of the metric, making it necessary to specify gauge conditions that fix the coordinate freedom. In the present four-dimensional formulation, the gauge conditions are  imposed directly on the spacetime metric, independently of the Einstein equations themselves.

A particularly natural choice for a fully covariant second-order formulation is the harmonic gauge, in which each coordinate function satisfies the harmonic condition
\begin{equation}
\Box x^a=-\Gamma^a=0
\quad\Longrightarrow\quad
\partial_b \left(\sqrt{-g}\,g^{ab}\right)=0.
\label{eq}
\end{equation}
This choice is especially appealing because, when $\Gamma^a=0$, the principal part of the Einstein equations is determined solely by the covariant wave operator acting on the metric components. As a result, the linearized system is manifestly hyperbolic and the associated initial value problem is well posed.

Our implementation is, however, not restricted to harmonic coordinates. Instead, we treat the gauge conditions as an additional set of equations,
\begin{equation}
\mathcal{C}_a[g]=0,
\label{eq:gauge}
\end{equation}
whose explicit form can be adapted to the particular spacetime under consideration. This separation between the field equations and the gauge conditions provides the flexibility to employ different coordinate choices for the benchmark problems considered in this work. In particular, some tests adopt a gauge in which the spatial coordinates are fixed by imposing $g^{0i}=0$, while the time coordinate remains harmonic.

%%%%%%%%%%%%%%%%%%%%%%%%%%%%%%%%%%%%%%
\subsection{Klein-Gordon equations}
%%%%%%%%%%%%%%%%%%%%%%%%%%%%%%%%%%%%%%%%

The dynamics of the complex scalar field $\Phi=\Phi_R+i\Phi_I$ is governed by the Klein--Gordon equation,
\begin{equation}
    \nabla_a\nabla^a\Phi =  \frac{dV}{d|\Phi|^2}\Phi ~~.
    \label{eq:kg}
\end{equation}
where $\nabla_a$ denotes the covariant derivative and $V(|\Phi|^2)$ is the scalar-field self-interaction potential. For the solitonic boson star evolutions presented in Sec.~\ref{sec:boson_star}, we employ the potential
\begin{eqnarray}
    V\left(|\Phi|^2\right) &=&
    m_b^2 |\Phi|^2
    \left(
        1-\frac{2|\Phi|^2}{\sigma_0^2}
    \right)^2,  \label{eq:solitonic_potential} \\
    \frac{dV}{d|\Phi|^2} &=& m_b^2
    \left(
        1-\frac{2|\Phi|^2}{\sigma_0^2}
    \right)
    \left(
        1-\frac{6|\Phi|^2}{\sigma_0^2}
    \right) \nonumber
\end{eqnarray}
%\eqref{eq:solitonic_potential}
where $m_b$ sets the boson mass scale and $\sigma_0$ controls the self-interaction strength. The Klein--Gordon equation takes the explicit form
\begin{equation}
\begin{split}
    \frac{1}{\sqrt{-g}}
    \partial_a
    \left(
        \sqrt{-g}\,g^{ab}\partial_b \Phi
    \right)
     =  \frac{dV}{d|\Phi|^2}\Phi
\end{split}
\label{eq:kg_real}
\end{equation}

The scalar field acts as the source of the spacetime curvature through its stress--energy tensor, which, for a complex scalar field, is given by
\begin{eqnarray}
    T_{ab}
    &=&
    \nabla_a\Phi\,\nabla_b\Phi^\ast 
    + \nabla_a\Phi^\ast\,\nabla_b\Phi 
    \\
    &-& g_{ab}
    \left[
        \nabla^c\Phi\,\nabla_c\Phi^\ast
        +
        V\left(|\Phi|^2\right)
    \right].
    \label{eq:stress_energy}
\end{eqnarray}

The complex scalar field action is invariant under global \(U(1)\) phase transformations, implying through Noether's theorem that there is a conserved current
\begin{equation}
    j^a
    =
    \frac{i}{2}
    \left(
        \Phi^\ast\nabla^a\Phi
        -
        \Phi\nabla^a\Phi^\ast
    \right),
    \label{eq:noether_current}
\end{equation}
The corresponding conserved Noether charge is obtained by integrating the time component of this current over a spacelike hypersurface. It can be interpreted as the conserved boson number and, for the boson star evolutions presented in Sec.~\ref{sec:boson_star}, provides an important global diagnostic of the accuracy of the evolution.

%%%%%%%%%%%%%%%%%%%%%%%%%%%%%%%%%%%%%%%%%%%%%%%%%%%%%%%%%5
\subsection{Residuals of the EKG system}
%%%%%%%%%%%%%%%%%%%%%%%%%%%%%%%%%%%%%%%%%%%%%%%%%%%%%%%%%%

The dynamical variables of the system, i.e., the outputs of the PINN, are the ten independent components of the spacetime metric in contravariant form, together with the real and imaginary parts of the complex scalar field, namely $(g^{ab},\Phi_R,\Phi_I)$. The corresponding field equations can be expressed in terms of residuals that vanish when the Einstein--Klein--Gordon equations and the chosen gauge conditions are satisfied. These residuals are functionals of the 12 dynamical fields and can be written schematically as
\begin{eqnarray}
\mathcal{E}_{ab}
&\equiv&
R_{ab}
- 8\pi
\left(
T_{ab}
- \frac{1}{2}T g_{ab}
\right),
\label{eq:einstein_residual}
\\
\mathcal{K}
&\equiv&
\frac{1}{\sqrt{-g}}
\partial_a
\left(
\sqrt{-g} g^{ab}\partial_b \Phi
\right)
- \frac{dV}{d|\Phi|^2}\Phi,
\label{eq:kg_residual}
\\
\mathcal{G}_a
&\equiv&
\mathcal{C}_a[g],
\label{eq:gauge_residual}
\end{eqnarray}
where $\mathcal{E}_{ab}$ represents the ten independent Einstein equations, $\mathcal{K}$ represents the complex Klein--Gordon equation and therefore two real equations, and $\mathcal{G}_a$ represents the four gauge conditions. Together, these provide the 16 residual equations enforced during PINN training.

A central aspect of our formulation is that the field equations are evaluated directly in their fully covariant four-dimensional form. In particular, we do not perform projections parallel and perpendicular to the spatial hypersurfaces to construct the evolution equations, nor do we introduce auxiliary variables to reduce the system to first order. The same covariant EKG equations are therefore used throughout, with only the coordinate conditions being modified according to the problem under consideration. The incorporation of the residuals given by Eqs.~(\ref{eq:einstein_residual}--\ref{eq:gauge_residual}) into the PINN optimization problem is described in Sec.~\ref{sec:pinns}.

%%%%%%%%%%%%%%%%%%%%%%%%%%%%%%%%%%%%%%%%%%%%%%%%%%%%%%%%%%%%%
%%%%%%%%%%%%%%%%%%%%%%%%%%%%%%%%%%%%%%%%%%%%%%%%%%%%%%%%%%%%%
\section{Physics-Informed Neural Networks }
\label{sec:pinns}
%%%%%%%%%%%%%%%%%%%%%%%%%%%%%%%%%%%%%%%%%%%%%%%%%%%%%%%%%%%%%
%%%%%%%%%%%%%%%%%%%%%%%%%%%%%%%%%%%%%%%%%%%%%%%%%%%%%%%%%%%%%

The goal of PINNs is to approximate solutions of differential equations directly using neural networks. The loss function in this case is constructed from the residuals of the governing equations, together with those associated with the prescribed initial and boundary conditions. These residuals are evaluated at a set of collocation points distributed over the corresponding domains for each type of constraint. Additional residual terms can also be included when the solution is known at a discrete set of points, for example, to incorporate observational or numerical data. Such terms, however, are not necessary for the present approach and are therefore not considered here. The parameters of the neural network are then adjusted to minimize the resulting loss function (i.e., a combination of the residuals), turning the solution of the differential equations into a nonlinear optimization problem.

In this section, we briefly describe the PINN framework used throughout this work, focusing on the aspects relevant to our implementation of the EKG system. Our aim is to develop a general formulation that can be applied consistently across all the test cases considered here, rather than tailoring the network to each individual problem. The loss function is constructed directly from the full EKG equations, without introducing any simplifications, allowing the same framework to be applied across different spatial dimensions. 

Within this common framework, the main differences between the test cases arise from the dimensionality of the problem and the distribution of the training points. In the one-dimensional tests, the additional spatial coordinates are simply omitted by setting the corresponding inputs to zero, while retaining the same basic network architecture. In the three-dimensional case, the network is instead trained using a more targeted distribution of points around the compact object. The network architecture, sampling strategy, and construction of the loss function are described in the following subsections, with further technical details of the implementation provided in the appendices.

%%%%%%%%%%%%%%%%%%%%%%%%%%%%%%
\subsection{Neural Network representation}
\label{subsec:NN_representation}
%%%%%%%%%%%%%%%%%%%%%%%%%%%%%%

The solutions of the equations of motion considered here are represented by a fully connected neural network (FCNN) \cite{fully}. Here, the network receives the vector of variables $\mathbf{x}$ which describes the solution in a given domain, that is,
\begin{equation}
\mathbf{a}^{(0)}=\mathbf{x} ~,
\end{equation}
where in the present application, this vector contains the spacetime coordinates $(t,x,y,z)$, as well as possibly additional parameters (e.g., some coefficients in the initial and boundary conditions) characterizing the solution. The vector $\mathbf{x}$ corresponds to the \emph{input layer}, where each of its entries represents a \emph{neuron} in this layer.

The input is then propagated through a sequence of \emph{hidden layers}, whose activations are defined recursively by
\begin{equation}
    \mathbf{a}^{(\ell)}
    =
    \sigma\left(
        \mathbf{W}^{(\ell)}\mathbf{a}^{(\ell-1)}
        +
        \mathbf{b}^{(\ell)}
    \right),
    \quad
    \ell=1,\ldots,L-1,
    \label{eq:hidden_layers}
\end{equation}
where $\mathbf{a}^{(\ell)}\in\mathbb{R}^{n_\ell}$, so $n_\ell$ denotes the number of neurons in layer $\ell$. Hence, in an FCNN a neuron of a given layer is \emph{connected} to (or, in other words, is a function of) the neurons of the previous layer.

On the other hand, $\mathbf{W}^{(\ell)}\in\mathbb{R}^{n_\ell\times n_{\ell-1}}$ and $\mathbf{b}^{(\ell)}\in\mathbb{R}^{n_\ell}$ are the corresponding \emph{weight}
matrix and \emph{bias} vector, respectively, and $\sigma$ is a piecewise nonlinear transformation, commonly known as the \emph{activation} function. The choice of this function determines the nonlinearity applied at each hidden layer. In our preliminary experiments, the hyperbolic tangent, $\sigma(u)=\tanh(u)$, consistently provided the most accurate results among the activation functions considered, and is therefore adopted throughout this work \cite{activacion,fully}.

Finally, the output layer maps the result of the last hidden layer to the predicted solution by simply considering
\begin{equation}
    \widehat{\mathbf{u}}_{\theta}(\mathbf{x})
    =
    \mathbf{W}^{(L)}\mathbf{a}^{(L-1)}
    +
    \mathbf{b}^{(L)},
    \label{eq:network_output}
\end{equation}
where $\widehat{\mathbf{u}}_{\theta}$ denotes the PINN approximation to the solution and $\theta$ represents the complete set of parameters that comprises all weight matrices and bias vectors. The output of the neural network, and therefore the quality of the corresponding PINN approximation, will then depend on the values of these parameters. 

Considering FCNNs as function approximators is mathematically well-motivated due to the Universal Approximation Theorem \cite{teorema,teorema2,teorema3}, which, under very mild conditions, states that they can \emph{uniformly} approximate any continuous function defined in a compact domain, given a sufficient number of parameters in $\theta$. Since PDE solutions are assumed to be differentiable, they can in theory be approximated by FCNNs.

However, it should be noted that this theorem does not specify either the number of parameters required or the particular set of parameters $\theta$ needed to obtain an accurate approximation. Thus, given a predefined number of hidden layers and neurons, the parameters in $\theta$ are adjusted iteratively so that the resulting neural network progressively improves its approximation of the desired function. This procedure is commonly referred to as the training process, and $\theta$ is therefore referred to as the set of trainable parameters.

%%%%%%%%%%%%%%%%%%%%%%%%%%%%%%
\subsection{Loss function}
\label{subsec:loss_function}
%%%%%%%%%%%%%%%%%%%%%%%%%%%%%%

To adjust the trainable variables, we need a criterion that measures the discrepancy between the desired function and its corresponding neural network approximation. This criterion is given by the \emph{loss function}. In the following, we present the formulation of the loss function used in the PINN framework.

Consider the following generic system of PDEs
\begin{equation}
    \mathcal{E}\left[\mathbf{u}(x^\mu)\right]=0,
\end{equation}
where $\mathbf{u}$ denotes the set of unknown fields, which depend on the coordinates $x^\mu$ defined in some domain, and $\mathcal{E}$ is a given nonlinear differential operator. If we replace the exact solution $\mathbf{u}$ by the corresponding PINN approximation $\widehat{\mathbf{u}}_{\theta}$ we obtain the residual
\begin{equation}
    \mathbf{r}_{\theta}(x^\mu)
    =
    \mathcal{E}\left[
        \widehat{\mathbf{u}}_{\theta}(x^\mu)
    \right] ~,
    \label{eq:pinn_residual}
\end{equation}
which measures the local violation of the governing equations at a given point in the input domain.

We can use these residuals to obtain a global measure of how closely the PINN approximation satisfies the PDE throughout the solution domain. To do that, we evaluate them at a given set of $N_c$ \emph{collocation} points. Squaring the residual yields a non-negative quantity, which is then averaged over all collocation points to define the PDE loss,
\begin{equation}
    \mathcal{L}_{\mathrm{PDE}}
    =
    \frac{1}{N_c}
    \sum_{i=1}^{N_c}    
        |\mathbf{r}_{\theta}(x_i^\mu)|^2.
    \label{eq:pde_loss}
\end{equation}
Thus, smaller values of $\mathcal{L}_{\mathrm{PDE}}$ indicate that the neural network approximation satisfies the governing equations more accurately, on average, over the region sampled by the collocation points.

Evaluating these residuals would be very costly if they were to be calculated by, for example, discretizing the domain in cells to compute a numerical approximation of the derivatives involved. However, in modern deep-learning frameworks, all operations to calculate the output of the neural network are recorded in \emph{graphs}. This allows these derivatives to be calculated for every collocation point very efficiently and to machine accuracy by inverting the recorded operations, a technique called Automatic Differentiation (AD) \cite{ad}. As a result, the collocation points do not need to be structured on any regular grid, and so they can even be resampled during the training process, as discussed below in Sec.~\ref{subsec:multiwindow}. 

On the other hand, the PINN approximation must also satisfy the initial and boundary conditions of the PDE problem under consideration. A possible way to incorporate them during the training process is through additional loss terms that penalize their violation consistently with the PDE system \cite{karniadakis2021physics}. This approach is known as \emph{soft-enforcement}. Denoting the corresponding contributions from the initial and boundary conditions by $\mathcal{L}_{\mathrm{ID}}$ and $\mathcal{L}_{\mathrm{BC}}$, respectively, the loss function considered for the training process can be written as
\begin{equation}
    \mathcal{L}_{\mathrm{tot}}
    =
    \lambda_{\mathrm{PDE}}\mathcal{L}_{\mathrm{PDE}}
    +
    \lambda_{\mathrm{ID}}\mathcal{L}_{\mathrm{ID}}
    +
    \lambda_{\mathrm{BC}}\mathcal{L}_{\mathrm{BC}},
    \label{eq:generic_total_loss}
\end{equation}
where $\lambda_{\mathrm{PDE}}$, $\lambda_{\mathrm{ID}}$, and $\lambda_{\mathrm{BC}}$ are weighting coefficients that determine the relative importance of the different loss terms. 

Another possibility is to incorporate them directly into the network ansatz so that they are satisfied identically for any value of the trainable parameters \cite{urban2023modelling}, a strategy known as \emph{hard-enforcement}. In our experience, soft-enforcement provides better results for the problems considered in this paper. We therefore adopt this approach throughout the main analysis of Sec.~\eqref{sec:tests}, while discussing hard enforcement separately below, where it is used only for the boundary conditions of some of the benchmark problems.

%%%%%%%%%%%%%%%%%%%%%%%%%%%%%%%%%%%%%%%%%%%%%%%.
\subsection{Training process}
%%%%%%%%%%%%%%%%%%%%%%%%%%%%%%%%%%%%%%%%%%%%%%%

The training process of a PINN therefore leads to an optimization problem, in which the parameters $\theta$ are iteratively adjusted to minimize the loss function in Eq.~\ref{eq:generic_total_loss}. The objective is to determine the set of parameters $\theta^\star$ such that,
\begin{equation}
\theta^\star = \underset{\theta}{\operatorname{arg \,min}}\,
\mathcal{L}(\theta).
\label{eq:nn_eq}
\end{equation}
Starting from an initial set of parameters, typically obtained through random initialization, an optimization algorithm (\emph{optimizer}) iteratively adjusts this set to reduce $\mathcal{L}$ in every iteration, with a correction proportional to the gradient of the loss function \cite{nocedal2006numerical}. To calculate the derivatives of the loss function with respect to the trainable variables we use again AD, allowing us to calculate the optimizer corrections at each step with machine precision.

At a stationary point of $\mathcal{L}$, this gradient vanishes and, consequently, no further correction is obtained. However, reaching such a point does not necessarily imply that we have found a \emph{global} minimum, since the loss function can be non-convex and contain multiple local minima and saddle points. The success of the optimization process therefore relies on the ability of the optimizer to efficiently reach a solution $\theta^*$ close to a global minimum of Eq.~\ref{eq:generic_total_loss}. 

Another important aspect is the \emph{rate of convergence} of these optimizers. In this respect, although the Adam optimizer \cite{2015-kingma} is widely used in the PINN literature, we find that quasi-Newton methods \cite{SSBRO} converge considerably faster and so achieve much higher accuracies for the problems considered here. We therefore employ throughout this work the self-scaled Broyden (\emph{SSBroyden}) quasi-Newton optimizer introduced in Ref.~\cite{urban2025unveiling}. This choice represents one of the main methodological differences between our approach and many previous PINN studies and has a significant impact on the accuracy of the resulting solutions. Further details of SSBroyden are provided in Appendix~\ref{app:optimization}.

%%%%%%%%%%%%%%%%%%%%%%%%%%%%%%%%%%%%%%%%%%%%%%%
\subsection{Setting the collocation points: multi-window,  resampling, and causality enforcement}
\label{subsec:multiwindow}
%%%%%%%%%%%%%%%%%%%%%%%%%%%%%%%%%%%%%%%%%%%%%%%

In the usual PINN formulation for time-dependent problems, a single neural network approximates the solution simultaneously in space and time over the entire domain of interest. For long-time evolutions, the network must therefore represent an increasingly large spacetime domain within a single approximation, which can make accurate training progressively more difficult. Moreover, minimizing the PDE and initial-data residuals over the full temporal interval does not explicitly enforce the causal structure of an initial-value problem: the optimizer can reduce the residual at later times before the solution at earlier times has been accurately determined. Different strategies have therefore been proposed to incorporate the causal structure of the evolution into the training procedure~\cite{causal_exp,multi}.

In the causal formulation of Ref.~\cite{causal_exp}, the PDE loss at each time is weighted by the loss accumulated at earlier times. Thus, later times receive less weight until the solution at earlier times is sufficiently accurate, encouraging the network to learn the solution progressively forward in time.
Alternatively, Ref.~\cite{multi} proposed a backward-compatible PINN, in which the same network is trained sequentially over successive time segments while preserving the solution learned in previous segments.

In this work, we combine some of these ideas, starting from the approach proposed in Ref.~\cite{multi}. We divide the full temporal domain into $N_w$ consecutive windows as:
\begin{equation}
    [t_0,T]
    =
    \bigcup_{k=0}^{N_w-1}
    [t_k,t_{k+1}],
    \qquad
    t_{N_w}=T,
    \label{eq:time_window_decomposition}
\end{equation}
and then train a separate neural network on each time window,
\begin{equation}
    \widehat{\mathbf{u}}_{\theta_k}^{(k)}
    (t,\mathbf{x}),
    \qquad
    t\in[t_k,t_{k+1}].
    \label{eq:window_network}
\end{equation}
The first network is constrained by the prescribed physical initial data. Once window $k$ has been trained, its prediction at the final time is used to provide the initial data for the subsequent window,
\begin{equation}
    \mathbf{u}_{\mathrm{ID}}^{(k+1)}(\mathbf{x})
    =
    \widehat{\mathbf{u}}_{\theta_k}^{(k)}
    (t_{k+1},\mathbf{x}).
    \label{eq:window_transfer}
\end{equation}
When first time derivatives are required as independent initial data, they are transferred as well. Since the solution in each window is represented by a smooth neural network, the fields and their derivatives at the end of the preceding window can be evaluated directly at the collocation points of the new initial surface, without requiring any interpolation.

This sequential procedure reduces the temporal extent represented by each network and introduces a causal structure by advancing the solution through successive time windows. It does not, however, eliminate error accumulation. Because the interface data are imposed only softly, inaccuracies in the solution at the end of one window may be propagated to the next and potentially amplified over successive intervals.

Despite the use of time windows, employing a very large number of collocation points within each window remains computationally expensive. We therefore use a fixed, computationally manageable set of collocation points that is periodically resampled during training. Although only a moderate number of points is evaluated at each optimization step, resampling progressively exposes the network to a much larger effective set of spacetime locations. This improves the coverage of each time window without increasing the computational cost of each step and reduces the tendency of the network to overfit a fixed set of points.

For some evolutions, the temporal distribution of the collocation points follows a time-dependent exponential law, initially concentrating points near the beginning of each window and progressively shifting the sampling toward later times. As training proceeds, the distribution gradually approaches uniform sampling over the full temporal interval. This provides a soft temporal bias toward earlier times, which is complemented by the causal weighting scheme of Ref.~\cite{causal_exp}. The latter prevents the optimizer from reducing the loss at later times while significant violations remain at earlier times, consistent with the forward-in-time propagation of the underlying hyperbolic system.

Within each window, we divide $[t_k,t_{k+1}]$ into $M$ temporal buckets of width $\Delta t = (t_{k+1}-t_k)/M$ and assign each collocation point $(t_i,\mathbf{x}_i)$ to a bucket
\begin{equation}
m_i =
\min\left(
\left\lfloor\frac{t_i-t_k}{\Delta t}\right\rfloor,
M-1
\right).
\end{equation}
Let $\overline{\mathcal L}_m$ denote the mean loss over the samples in bucket $m$, with the gradient not propagated through this quantity.\footnote{That is, although $\overline{\mathcal L}_m$ depends on the network parameters, it is treated as a constant when computing gradients, so no gradient is propagated through this dependence.} Each point is then assigned the weight
\begin{equation}
w_i =
\exp\left(
-\epsilon_c
\sum_{n=0}^{m_i-1}\overline{\mathcal L}_n
\right),
\end{equation}
where $\epsilon_c$ controls the strength of the causal weighting. The loss within the window is then given by
\begin{equation}
    \mathcal L =
    \frac{1}{N_c}\sum_{i=1}^{N_c} w_i\mathcal L_i.
\end{equation}
The contribution from the solution at the final time of each window is weighted analogously, using the accumulated mean loss over all $M$ buckets.

Finally, all coordinates supplied to the network are rescaled to the interval $[-1,1]$. This normalization avoids large input values that can drive the hyperbolic tangent activation function into its saturation regime, where its sensitivity to the input is reduced. It also puts the different coordinates on comparable scales, improving the conditioning of the optimization. In particular, the temporal coordinate is normalized independently within each time window according to
\begin{equation}
    \tau_k  =   2\frac{t-t_k}{t_{k+1}-t_k}-1 ~,
    \quad
    \tau_k\in[-1,1].
    \label{eq:local_time_scaling}
\end{equation}
This local transformation maps the entire duration of each window onto the same interval $[-1,1]$. As a result, the temporal dependence within an individual window does not become increasingly compressed as the total evolution time grows, and each network operates over a comparable range of its $\tanh$ activation function.

%%%%%%%%%%%%%%%%%%%%%%%%%%%%%%%%%%%%%%%%%%%%%%%%%
\subsection{Periodic boundary conditions through hard enforcement}
\label{subsec:conditions}
%%%%%%%%%%%%%%%%%%%%%%%%%%%%%%%%%%%%%%%%%%%%%%%%%

Initial and boundary conditions can also be imposed through hard enforcement by constructing the network ansatz such that the corresponding conditions are satisfied identically, independently of the values of the trainable parameters. A detailed discussion of this approach can be found in \cite{cc_periodicas}.

The vacuum benchmark problems considered below are periodic in the spatial directions. We therefore enforce periodicity exactly by replacing each spatial coordinate $x$ in the network input with a pair of periodic features, $\left(\sin kx,\cos kx\right)$. For the fundamental mode, $k=2\pi/L$, and for $x\in[-L/2,L/2]$ the neural representation can be written as
\begin{equation}
    \widehat{\mathbf{u}}_{\theta}(t,x) =
    \widehat{\mathbf{u}}_{\theta}
    \left(
        t,
        \sin\left(\frac{2\pi x}{L}\right),
        \cos\left(\frac{2\pi x}{L}\right)
    \right).
    \label{eq:periodic_embedding}
\end{equation}
Since the embedded features are invariant under the transformation $x\rightarrow x+L$, the network prediction is periodic by construction,
\begin{equation}
    \widehat{\mathbf{u}}_{\theta}(t,x+L)
    =
    \widehat{\mathbf{u}}_{\theta}(t,x).
\end{equation}
Furthermore, since both the trigonometric embedding and the neural network are smooth functions, all spatial derivatives obtained through automatic differentiation are periodic as well. The same construction is applied independently to each periodic spatial coordinate. This eliminates the need for an explicit periodic boundary loss and avoids introducing competition between the periodicity constraint and the other contributions to the training objective.

A hard enforcement construction for initial conditions can alter the scaling of solution errors and the optimization properties of the neural ansatz away from the initial hypersurface. A comparison between hard and soft enforcement of the initial data is provided in Appendix~\ref{app:hard_soft}. 

%%%%%%%%%%%%%%%%%%%%%%%%%%%%%%%%%%%%%%%%%%%%%%%
%%%%%%%%%%%%%%%%%%%%%%%%%%%%%%%%%%%%%%%%%%%%%%%
\section{One-Dimensional Vacuum Tests: Apples with Apples}
\label{sec:tests}
%%%%%%%%%%%%%%%%%%%%%%%%%%%%%%%%%%%%%%%%%%%%%%%
%%%%%%%%%%%%%%%%%%%%%%%%%%%%%%%%%%%%%%%%%%%%%%%

We evaluate the PINN formulation using standard benchmarks in numerical relativity, commonly known as the Apples with Apples tests: the linear wave, gauge wave, and polarized Gowdy tests \cite{mexico,mexicoZ4}. All three are vacuum spacetimes, so the scalar field and its derivatives are set identically to zero, reducing the EKG implementation to the vacuum Einstein system. These tests allow us to compare the PINN solutions with the corresponding analytical solutions while enforcing the second-order Einstein equations and the selected gauge conditions directly through the training loss. The spatial coordinates are sampled uniformly, since the nontrivial variations of the geometry are distributed throughout the computational domain.

The linear and nonlinear gauge wave tests are used to assess long-term propagation on periodic domains. The Gowdy spacetime, in contrast, provides a more demanding benchmark due to the presence of widely separated scales. In this latter case, we consider both the expanding and collapsing regimes, corresponding, respectively, to evolution away from and toward the cosmological singularity.

Spatial periodicity is imposed through the periodic input described in Sec.~\ref{sec:pinns}, while long evolutions are performed using the sequential time-window decomposition. The exact solutions are used only to construct the initial data and to assess the accuracy of the trained models, but are not used during training. Details of the neural network architecture used for each model are provided in Appendix~\ref{app:modelos}.

%%%%%%%%%%%%%%%%%%%%%%%%%%%%%%%%%%%%%%%%%%%
\subsection{Linear wave test}
\label{sec:linear_wave}

We first consider the standard linear wave benchmark, consisting of a plane wave propagating along the $x$ direction on a periodic spatial torus. The 
analytical metric, for the harmonic gauge $\Gamma^{a}=0$, is given by
\begin{equation}
    g_{\mu\nu}
    =
    \mathrm{diag}
    \left(
        -1,\,
        1,\,
        1+b,\,
        1-b
    \right).
\end{equation}

where we have defined

\begin{equation}
    b(t,x)  =
    A\sin\left[\frac{2\pi(x-t)}{d}\right].
\end{equation}

The metric satisfies the Einstein equations only to linear order in the wave amplitude $A$, with corrections appearing at $\mathcal{O}(A^2)$. Since the PINN is trained on the full nonlinear Einstein equations, the analytical metric given above is therefore not an exact solution of the system solved by the network, but rather provides a linearized reference solution. For sufficiently small $A$, however, the $\mathcal{O}(A^2)$ corrections are negligible, making it an appropriate benchmark for assessing the accuracy of the PINN. We therefore adopt the small amplitude $A=10^{-5}$.
The spatial domain is $x\in[-1,1]$, such that $d=2$, and the evolution is carried out up to $t=100$, corresponding to 50 crossing times.

\begin{figure*}[t]
    \centering
    \includegraphics[width=0.8\textwidth]{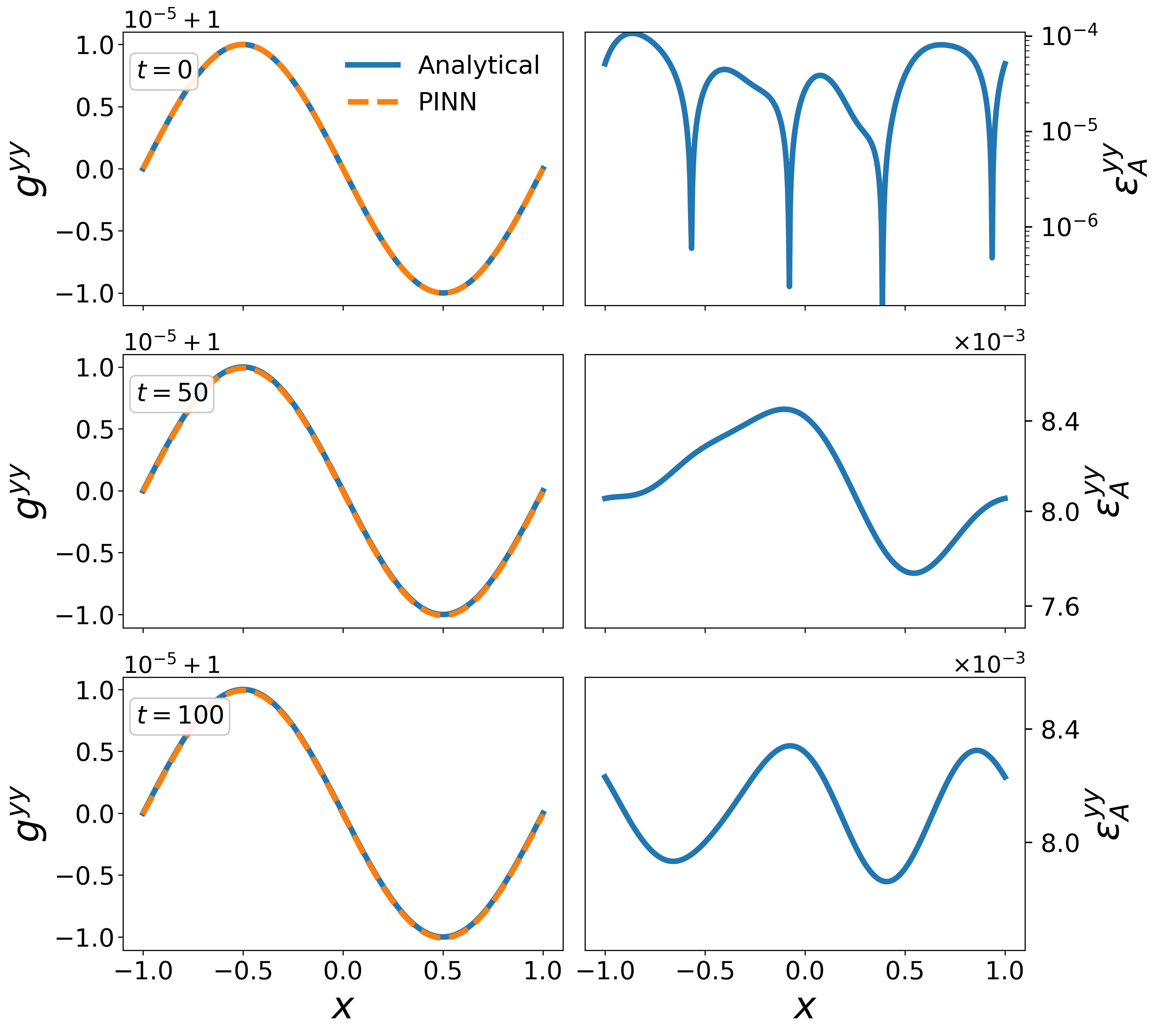}
    \caption{\emph{Linear wave test}. Comparison between the reference
    solution and the PINN prediction for the $g^{yy}$ component at
    representative times. The left panels show the analytical and
    PINN-predicted profiles, while the right panels display the absolute
    error normalized by the wave amplitude $A$, as defined in
    Eq.~\eqref{eq:linear_wave_normalized_error}.}
    \label{fig:lw_comparison}
\end{figure*}

Figure~\ref{fig:lw_comparison} compares the analytical and PINN solutions
for $g^{yy}$ at representative times. To quantify the local discrepancy relative to the characteristic scale of the
perturbation, we define the normalized  error\footnote{Normalizing by the constant amplitude $A$, rather than by the local
perturbation $|g_{\mathrm{exact}}^{yy}-1|$, avoids singular or artificially
large values at the nodes of the wave.}
\begin{equation}
    \epsilon_A^{yy}(t,x)
    =
    \frac{
        \left|
            g_{\mathrm{exact}}^{yy}(t,x)
            -
            g_{\mathrm{PINN}}^{yy}(t,x)
        \right|
    }{A}.
\label{eq:linear_wave_normalized_error}
\end{equation}    
As shown in the figure, the network accurately preserves both the sinusoidal profile and its phase throughout the evolution, with no visible deviations even after many crossing times 
\onecolumngrid
\twocolumngrid

\begin{figure}[H]
    \centering
    \includegraphics[width=1.0\linewidth]{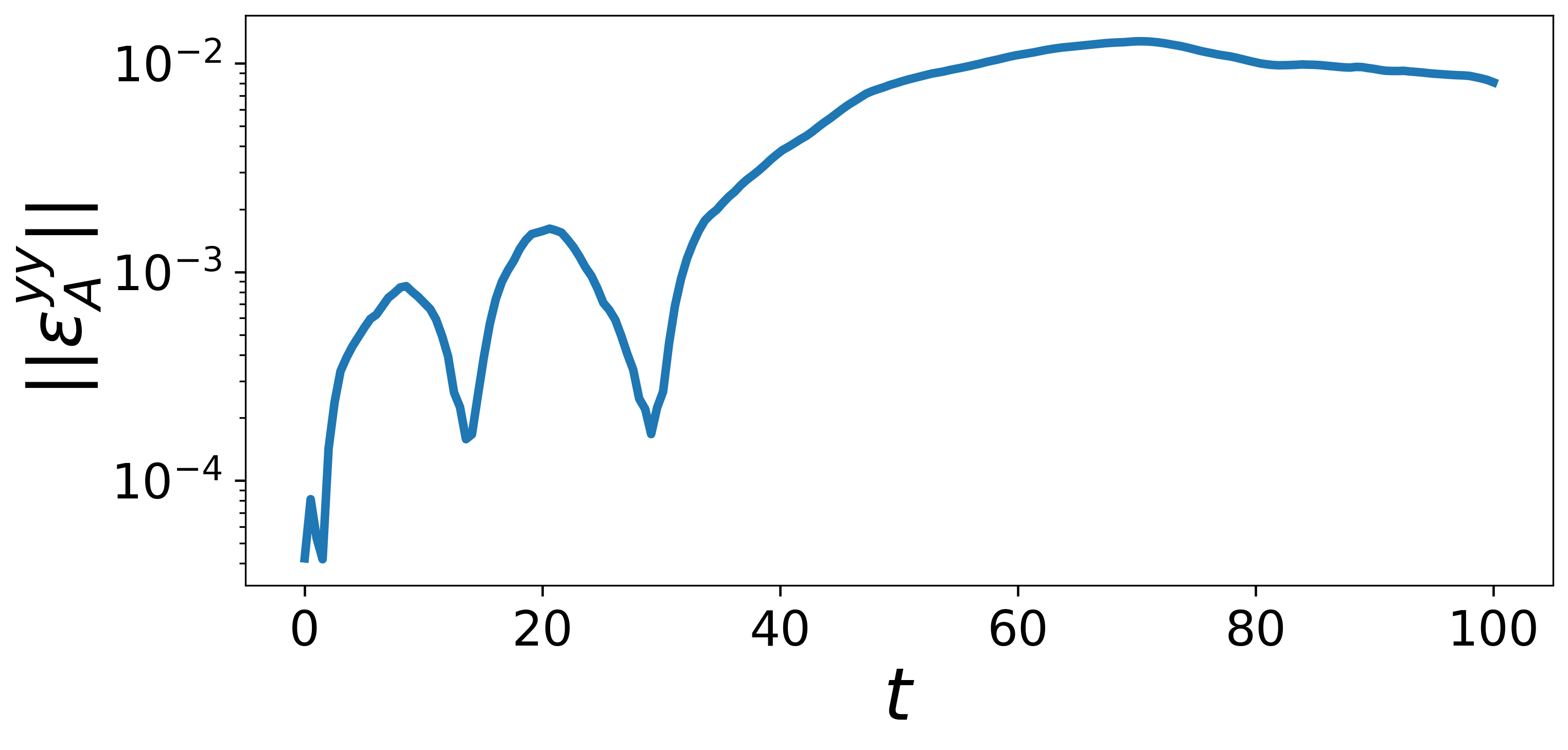}
    \caption{ {\em Linear wave test}.
        Temporal evolution of the spatially averaged absolute error normalized
        by the wave amplitude $\epsilon_A^{yy}$. After an initial
        growth, the error reaches values of order $10^{-2}$ and enters an
        approximately stationary regime. No sustained linear or exponential
        growth is observed over the $50$ crossing times considered.
    }
    \label{fig:lw_error_evolution}
\end{figure}

To quantify the error accumulation through the  evolution, we define the spatially averaged normalized error 
\begin{equation}
\begin{aligned}
||\epsilon_A^{yy}(t_k)||
&=
\frac{1}{N_x}
\sum_{i} \epsilon_A^{yy}(t_k,x_i),
\end{aligned}
\label{eq:linear_wave_mean_normalized_error}
\end{equation}
which is shown in Figure~\ref{fig:lw_error_evolution}. The error is evaluated on a uniform temporal grid over $t\in[0,100]$. At each time $t_k$, the PINN corresponding to the relevant time window is evaluated at $N_x=500$ uniformly spaced points along the $x$ direction, with the transverse coordinates fixed at $y=z=0$. These evaluation points are independent of the collocation points used during training.
The error initially increases before reaching values of order $10^{-2}$ at $t\approx 50$, after which it remains approximately constant throughout the remainder of the evolution.

%%%%%%%%%%%%%%%%%%%%%%%%%%%%%%%%%%%%%%%%%%
\subsection{Gauge wave test}
\label{subsec:gauge_wave}
%%%%%%%%%%%%%%%%%%%%%%%%%%%%%%%%%%%%%%%%%%

We next consider the standard gauge wave benchmark, which describes Minkowski spacetime in a nontrivial harmonic coordinate system satisfying $\Gamma^{a}=0$. For a wave propagating along the $x$ direction on a spatial three-torus, the exact spacetime metric is given by
\begin{equation}
    g_{\mu\nu}
    =
    \mathrm{diag}
    \left(
        -H,
        H,
        1,
        1
    \right),
    \label{eq:gauge_wave_metric}
\end{equation}
where
\begin{equation}
    H(t,x)
    =
    1
    -
    A\sin\left[
        \frac{2\pi(x-t)}{d}
    \right].
    \label{eq:gauge_wave_H}
\end{equation}

We set the amplitude to $A=0.1$, for which the nonlinear dependence of the metric on $H$ is no longer negligible. As before, we consider the spatial domain $x\in[-1,1]$. The solution is evolved up to $t=100$, corresponding to $50$ crossing times. Spatial periodicity is enforced through the periodic input embedding described in Sec.~\ref{sec:pinns}, while the long-time evolution is handled using the sequential time-window decomposition.

\begin{figure}[H]
    \centering
    \includegraphics[width=1.0\linewidth]{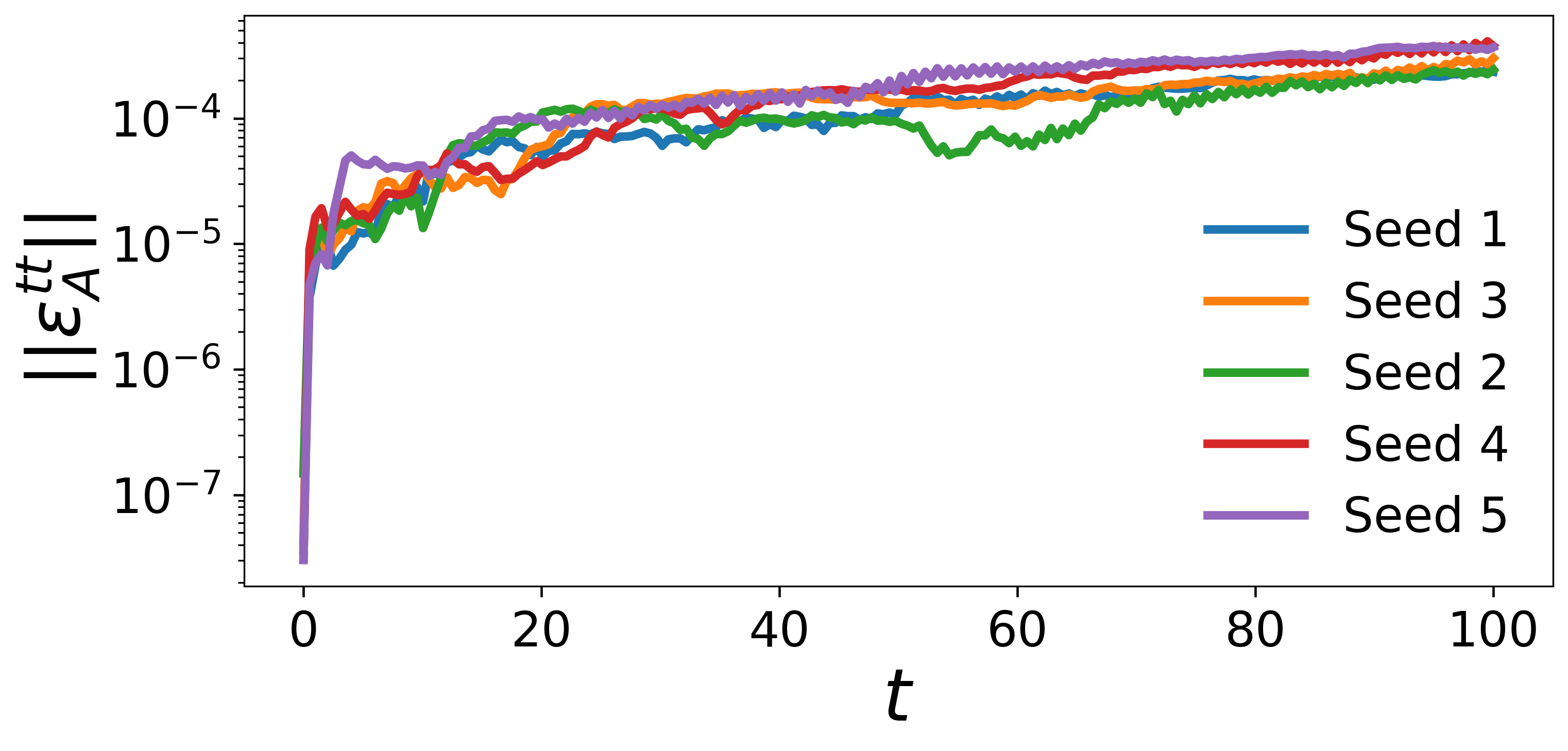}
    \caption{ {\em Gauge wave test}.
        Dependence of the gauge-wave evolution on the random seed. The curves show the spatially averaged absolute error in $g^{tt}$, normalized by the wave amplitude $A$, for five independent runs. All runs exhibit similar behavior, with errors remaining of $\mathcal{O}(10^{-4})$ over most of the evolution, indicating robustness against stochastic initialization and collocation sampling.  }
    \label{fig:gw_seeds}
\end{figure}

\begin{widetext}
    
\begin{figure}[H]
    \centering
    \includegraphics[width=0.8\linewidth]{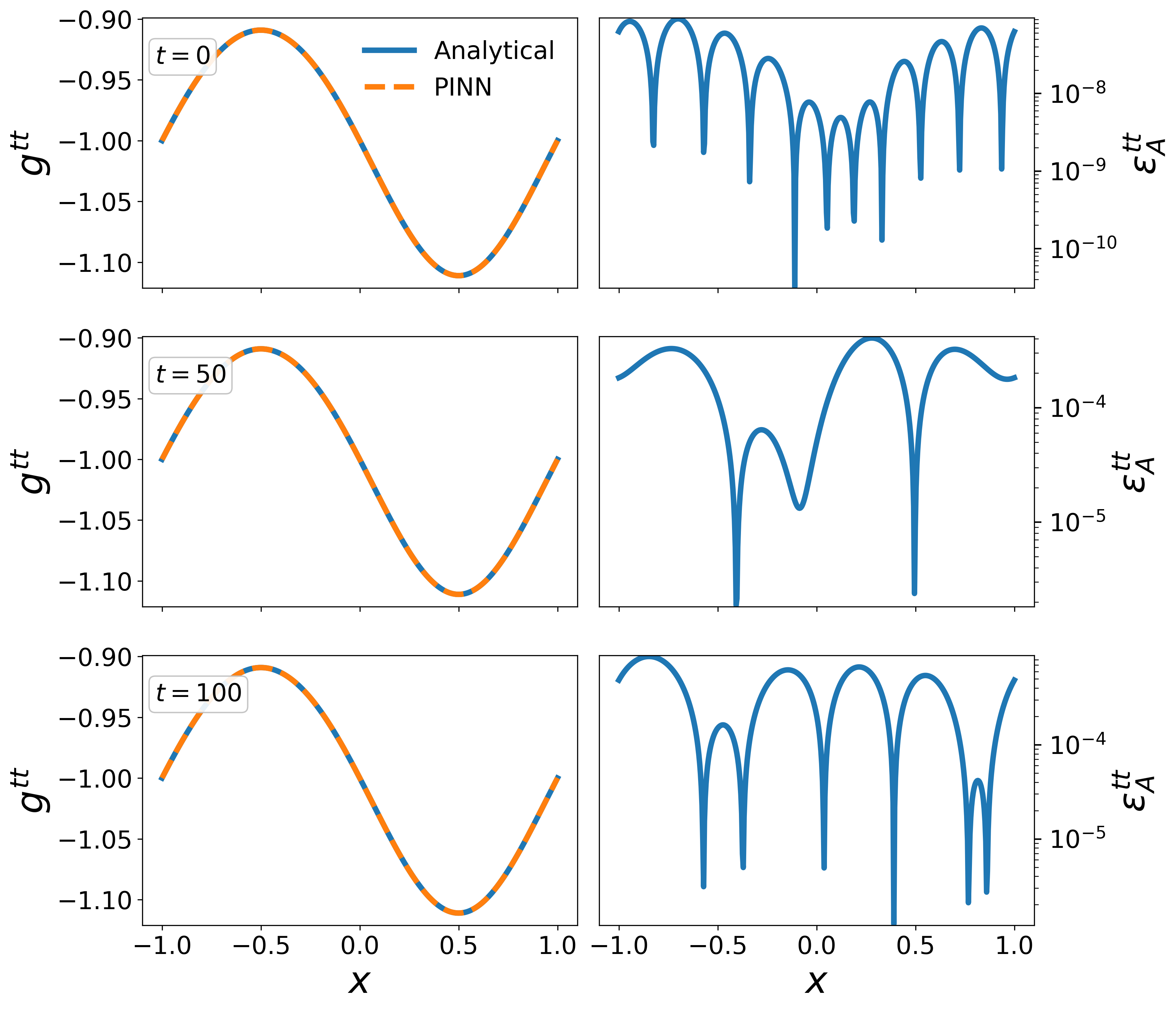}
    \caption{ {\em Gauge wave test}.
        Comparison between the exact solution and the PINN prediction for
        the $g^{tt}$ component at representative times.
        The left panels show the exact and predicted profiles, while the right
        panels display the absolute error normalized by the gauge wave
        amplitude $A$ on a logarithmic scale.
    }
    \label{fig:gw_comparison}
\end{figure}
\end{widetext}

Figure~\ref{fig:gw_comparison} compares the exact and PINN solutions at representative times, together with the normalized error for the $g^{tt}$ component, defined in Eq.~\ref{eq:linear_wave_normalized_error}. The network preserves both the spatial profile and the phase of the solution over repeated propagation across the periodic domain. To quantify the error accumulation, we use the spatially averaged normalized error defined in Eq.~\ref{eq:linear_wave_mean_normalized_error}, for the $g^{tt}$ component. As shown by the blue line in Fig.~\ref{fig:gw_seeds}, the error increases rapidly at early times, after which the error growth slows down and remains at approximately $\mathcal{O}(10^{-4})$ for most of the evolution.

We assess the sensitivity of the gauge wave evolution to the stochastic elements of the training procedure by repeating the same experiment with five independent random seeds. The seed affects both the initialization of the network parameters and the random sampling of collocation points during training. Figure~\ref{fig:gw_seeds} shows the evolution of the spatially averaged normalized error for the five runs. The solutions show good agreement across the different seeds, with moderate dispersion that becomes more noticeable at late times. This indicates that the gauge wave evolution is relatively robust to the stochastic variability associated with network initialization and collocation-point sampling.

\onecolumngrid
\twocolumngrid

\label{subsec:gauge_wave}

%%%%%%%%%%%%%%%%%%%%%%%%%%%%%%%%%%%%%%%%%%%%%
\subsubsection{Parametric gauge-wave family}
\label{subsec:parametric_gauge_wave}
%%%%%%%%%%%%%%%%%%%%%%%%%%%%%%%%%%%%%%%%%%%%%

A particularly attractive feature of the PINN formulation is that physical parameters can be incorporated directly as additional inputs to the network. This allows a single trained model to represent not just a particular solution of the Einstein equations, but a continuous family of solutions parameterized by the underlying physical parameters.

To illustrate this capability, we treat the gauge wave amplitude $A$ as an additional input parameter of the network,
\begin{equation}
\hat{\mathbf{u}}_\theta =
\hat{\mathbf{u}}_\theta(t,x_i;A).
\label{eq:parametric_gauge_network}
\end{equation}
Rather than training a separate network for each value of $A$, the network is trained to represent a family of solutions parametrized by the wave amplitude. During training, the amplitude is sampled uniformly in $\log_{10}A$ over the physical range $A\in[10^{-3},10^{-1}]$. Before being provided to the PINN, it is normalized to the interval $[-1,1]$ using the logarithm of the amplitude. 
%Thus, $A=10^{-3}$, $10^{-2}$, and $10^{-1}$ correspond to network inputs of $-1$, $0$, and $1$, respectively.
This normalization provides a well-scaled input across the full amplitude range while preserving the logarithmic sampling.
%
%\kw{During training, amplitudes are sampled uniformly in $\log_{10}A$ over the physicall range $A \in[10^{-3},10^{-1}]$. Before being passed to the PINN, the amplitude is transformed as:}
%\begin{equation}
%A_{\mathrm{net}} =
%2\,\frac{\log_{10}A-\log_{10}A_{\min}}
%{\log_{10}A_{\max}-\log_{10}A_{\min}}-1,
%\qquad
%\end{equation}
%\kw{Thus, $A=10^{-3},10^{-2},10^{-1}$ correspond to network inputs $-1,0,1$, respectively. This scaling places equal changes in $A$ at equal distances in the network input and keeps that input within $[-1,1]$, avoiding large input magnitudes that could saturate the activation functions.}

Once training is complete, the network can be evaluated at different values of $A$ within the training range, allowing different members of the solution family to be generated without retraining the network. 
This represents an important distinction from conventional numerical evolutions based on finite-difference or spectral discretizations. In such approaches, changing the physical parameter $A$ generally requires repeating the evolution for the corresponding initial data. By contrast, the dependence on $A$ is learned simultaneously within a single training procedure. The resulting PINN therefore provides a continuous parametric representation of the solution family, allowing the evolution for any amplitude within the training range to be evaluated directly after training.

\begin{figure}[H]
\centering
\includegraphics[width=\linewidth]{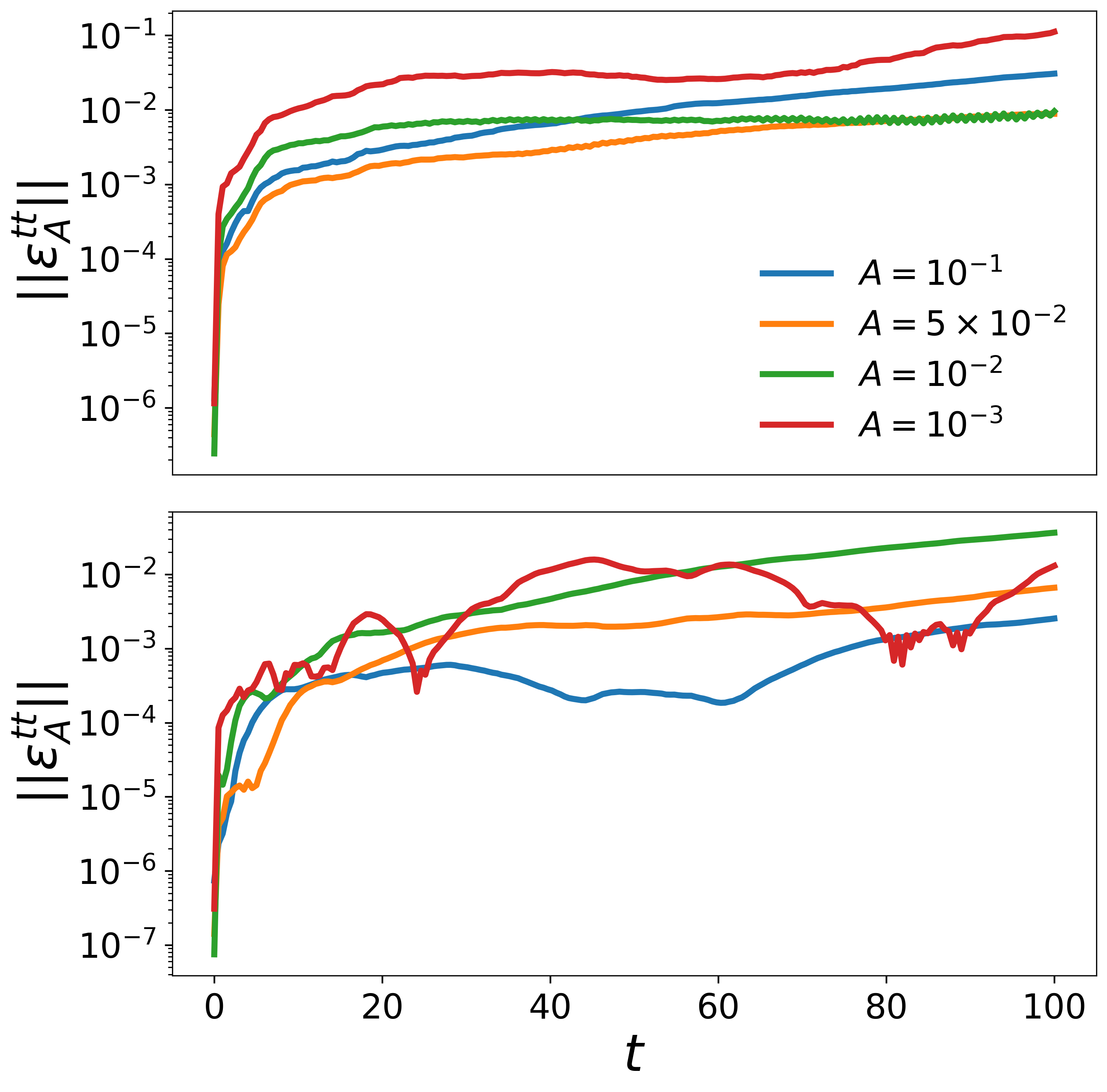}
\caption{ {\em Gauge wave test}.
Temporal evolution of the spatially averaged absolute error in
$g^{tt}$, normalized by the gauge-wave amplitude, for four representative values of $A$. (Top) Continuous
random sampling over $A\in[10^{-3},10^{-1}]$. (Bottom) 
The parameter-space boundaries are reinforced with additional collocation points. Each panel corresponds to a separately trained
parametric PINN, while all four amplitudes within each panel are predicted by the same network.}
\label{fig:gw_parametric_amplitudes}
\end{figure}

We assess the accuracy of the parametric representation by considering four representative amplitudes spanning the training interval. The top panel of Fig.~\ref{fig:gw_parametric_amplitudes} shows the normalized mean error in $g^{tt}$ as a function of time for these amplitudes, with the network trained using values of $A$ sampled continuously across the full allowed range. The errors for the intermediate amplitudes, $A=5\times10^{-2}$ and $A=10^{-2}$, remain close to or below $10^{-2}$ throughout the evolution, whereas larger errors are observed at the endpoints, $A_{min}=10^{-3}$ and $A_{max}=10^{-1}$.

This behavior suggests that the endpoints are less effectively constrained by the continuous sampling strategy. For an interior value of $A$, nearby training points are sampled on both sides of the target value, providing information about the local dependence of the solution on the parameter. In contrast, near $A_{\min}$ or $A_{\max}$, all sampled points lie on only one side of the boundary. Consequently, the effective neighborhood of training points available to constrain the solution is reduced near the endpoints, which can make the parametric representation less accurate there. To test whether this boundary effect contributes to the observed errors, we repeat the training while adding 200 collocation points at $A=A_{\min}$ and another 200 at $A=A_{\max}$. Their spacetime coordinates are sampled randomly in the same way as those of the remaining training points. The resulting errors are shown in the bottom panel of Fig.~\ref{fig:gw_parametric_amplitudes}.

The comparison between the two panels shows that explicitly sampling
$A_{\min}$ and $A_{\max}$ reduces the endpoint errors and produces a more
uniform accuracy across the parameter interval. This confirms that the
degradation observed in the initial training is primarily associated with the
weaker representation of the parameter-space boundaries. Importantly, the
additional samples do not require separate models for the limiting cases:
a single parametric PINN still represents the complete continuous family of
solutions.

%%%%%%%%%%%%%%%%%%%%%%%%%%%%%%%%%%%%%%%%%%%%%%%%%%
\subsubsection{Sensitivity to network capacity and collocation density}

We have performed a controlled study using the gauge-wave test, in which the network architecture and the number of collocation points are varied independently to assess the robustness of the results and determine how the accuracy depends on the numerical setup. The physical configuration, time-window decomposition, optimization procedure, and total number of training iterations are kept unchanged throughout the study. Although we do not repeat this analysis for all test cases, additional experiments in other configurations show similar trends, suggesting that the conclusions are not specific to the gauge-wave test.

\begin{figure}[H]
    \centering
    \includegraphics[width=0.45\textwidth]{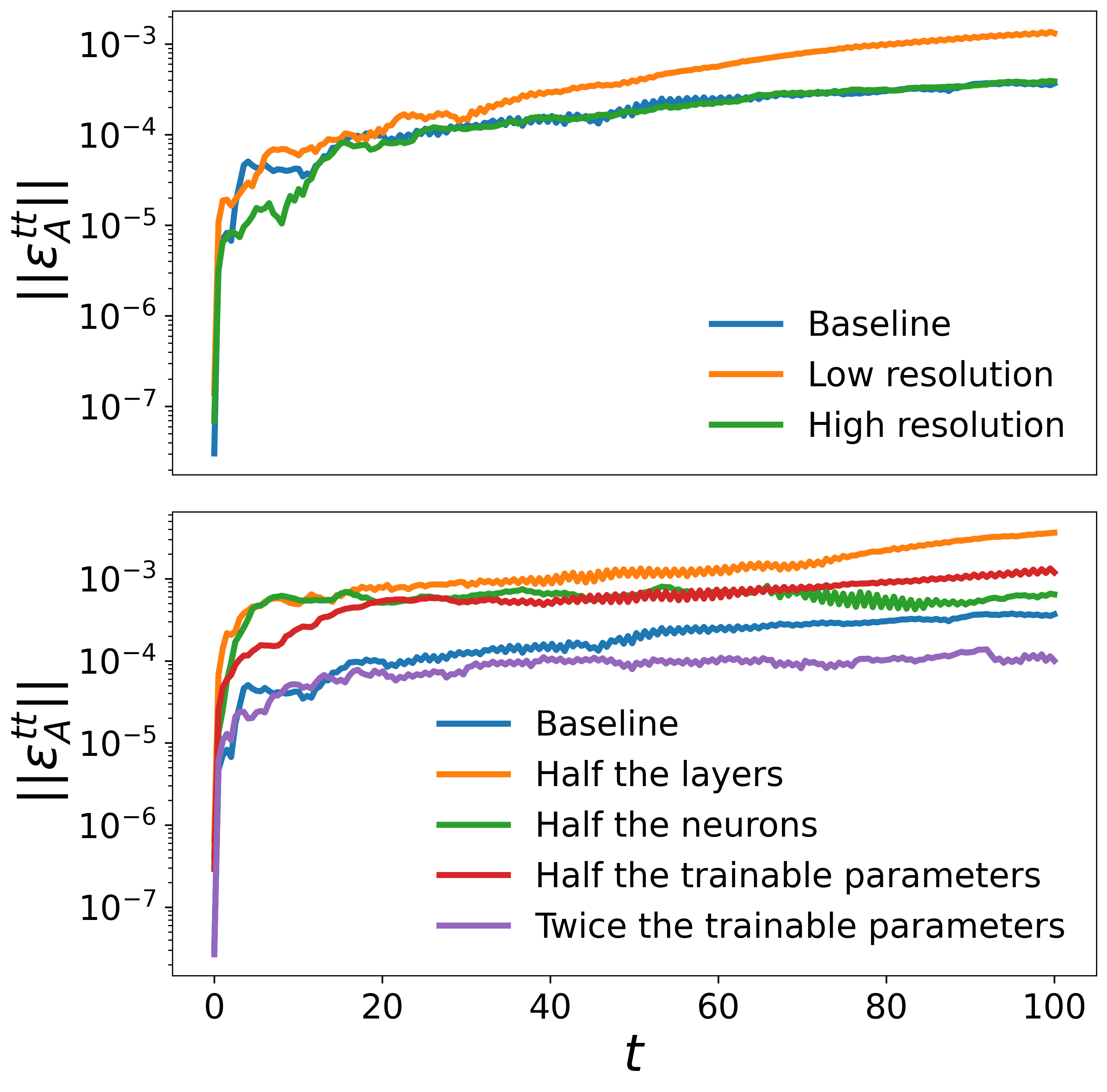}
    \caption{ {\em Gauge wave test}.(Top) Dependence on the number of collocation points. Reducing their number increases the error, while increasing it beyond the baseline value gives little improvement. (Bottom) Dependence on network capacity. Smaller networks lead to larger long-time errors, whereas increasing the number of trainable parameters gives the lowest error.}
    \label{fig:gauge_convergence}
\end{figure}

The top panel of Figure~\ref{fig:gauge_convergence} examines the dependence of the solution on the number of collocation points, which is varied by a factor of two relative to the baseline case. Reducing their number effectively lowers the resolution of the spacetime domain and leads to a clear loss of accuracy. The low-resolution configuration reaches errors of order $10^{-3}$ at late times, several times larger than those of the baseline simulation. These results indicate that sufficiently dense sampling of the spacetime domain is needed to constrain the equation residuals throughout each time window.

Increasing the number of collocation points beyond the baseline, however, produces little additional improvement. The high-resolution and baseline results remain close throughout most of the evolution and reach similar errors at the final time. This suggests that the baseline sampling is already sufficient for the chosen network architecture and training procedure. Further increasing the sampling density therefore does not significantly improve the accuracy, which appears to be limited by other factors, such as the network capacity, the training procedure, or error accumulation across successive time windows.

The bottom panel of Figure~\ref{fig:gauge_convergence} shows the effect of modifying the neural network architecture. Reducing the number of hidden layers produces the largest loss of accuracy: the error rapidly reaches values of order $10^{-3}$ and continues to grow throughout the evolution, reaching values several times larger than those of the baseline configuration. Reducing either the number of neurons per layer or the total number of trainable parameters also increases the error, although the effect is less pronounced. These results suggest that the baseline architecture is not strongly overparameterized and that sufficient network capacity is needed to accurately represent the nonlinear spacetime dependence across successive time windows.

Conversely, increasing the number of trainable parameters systematically improves the prediction. The corresponding error remains below the baseline curve over almost the entire evolution, reaching values of order $10^{-5}$ at the final time. The improvement is particularly evident at late times, when errors accumulated at successive time-window interfaces become more significant. This dependence on network size suggests that a substantial part of the error in the baseline evolution is related to the limited capacity of the network, rather than solely to the temporal decomposition.

We emphasize that these results should not be interpreted as a classical convergence test, since the accuracy of a PINN depends on both the network architecture and the training procedure. Nevertheless, they reveal a clear trend: reducing either the number of trainable parameters or the number of collocation points degrades the solution, while increasing the network size improves the long-term accuracy. In contrast, increasing the number of collocation points beyond the baseline has little effect, indicating that the baseline sampling is already sufficient for the chosen network and training procedure. %Further details on the neural network training times for these models are provided in Appendix~\ref{app:modelos}.

%%%%%%%%%%%%%%%%%%%%%%%%%%%%%%%%%%%%%%%%
\subsection{Gowdy spacetime test}
\label{sec:gowdy}
%%%%%%%%%%%%%%%%%%%%%%%%%%%%%%%%%%%%%%%%

A more demanding test involving a highly curved spacetime is provided by the polarized Gowdy solution~\cite{mexico,mexicoZ4}. The corresponding metric is given by
\begin{equation}
    g_{\mu\nu }  =
    \operatorname{diag}
    \left(
        -t^{-1/2}\mathrm{e}^{\lambda/2},
        t\mathrm{e}^{P},
        t\mathrm{e}^{-P},
        t^{-1/2}\mathrm{e}^{\lambda/2}
    \right),
    \label{eq:gowdy_metric}
\end{equation}
where $P=P(t,z)$ and $\lambda=\lambda(t,z)$ are periodic in the $z$ direction. For the standard benchmark considered here, an analytical solution is obtained by choosing
\begin{equation}
    P(t,z)  =   J_0(2\pi t)\cos(2\pi z),
    \label{eq:gowdy_P}
\end{equation}
with the corresponding function $\lambda(t,z)$ given by
\begin{align}
    \lambda(t,z)
    ={}&
    -2\pi t
    J_0(2\pi t)J_1(2\pi t)
    \cos^2(2\pi z)
    \nonumber\\
    &+
    2\pi^2t^2
    \left[
        J_0^2(2\pi t)
        +
        J_1^2(2\pi t)
    \right]
    \nonumber\\
    &-
    \frac{1}{2}
    \left\{
        (2\pi)^2
        \left[
            J_0^2(2\pi)
            +
            J_1^2(2\pi)
        \right]
    \right.
    \nonumber\\
    &\left.
        -
        2\pi J_0(2\pi)J_1(2\pi)
    \right\},
    \label{eq:gowdy_lambda}
\end{align}
with $J_0$ and $J_1$ denoting Bessel functions.

The solution has several properties that make it a challenging test. The spacetime is singular at $t=0$, corresponding to the cosmological singularity, while $\lambda$ evolves significantly during the expansion and enters exponentially into some of the metric components. As a result, the metric components can develop very different magnitudes and growth rates, so even relatively small errors in $\lambda$ can be amplified in the metric. The Gowdy test therefore probes the ability of the PINN to represent nonlinear spacetime dynamics across multiple scales.

As in the standard benchmark, we consider both directions of temporal evolution of the exact solution. The forward evolution describes the expanding cosmology and moves away from the singularity, whereas the collapsing evolution proceeds toward it. In the original benchmark, finite-difference simulations were reported to break down at approximately $t \simeq 4$ in the expanding case, while the collapsing evolution remained stable up to the formation of the singularity~\cite{mexico}.

%%%%%%%%%%%%%%%%%%%%%%%%%%%%%%%%%%%%%%%%%%%
\subsubsection{Expanding evolution}
\label{sec:gowdy_expansion}
%%%%%%%%%%%%%%%%%%%%%%%%%%%%%%%%%%%%%%%%%%

The evolution starts from the exact solution at $t=1$ and proceeds through successive time windows up to the final time $t=8$. We impose the gauge conditions used in the original Gowdy benchmark~\cite{mexico}. In terms of the covariant metric components, these conditions are
\begin{equation}
g_{ti}=0,
\qquad
g_{tt}=-g_{zz}.
\end{equation}
The initial data and the exact solution used for comparison with the PINN predictions are both constructed from Eqs.~\eqref{eq:gowdy_metric}--\eqref{eq:gowdy_lambda}. Throughout the evolution, we monitor the inverse metric components $g^{tt}$, $g^{yy}$, and $g^{zz}$.

\begin{widetext}
\begin{figure}[H]
    \centering
    \includegraphics[width=0.9\linewidth]{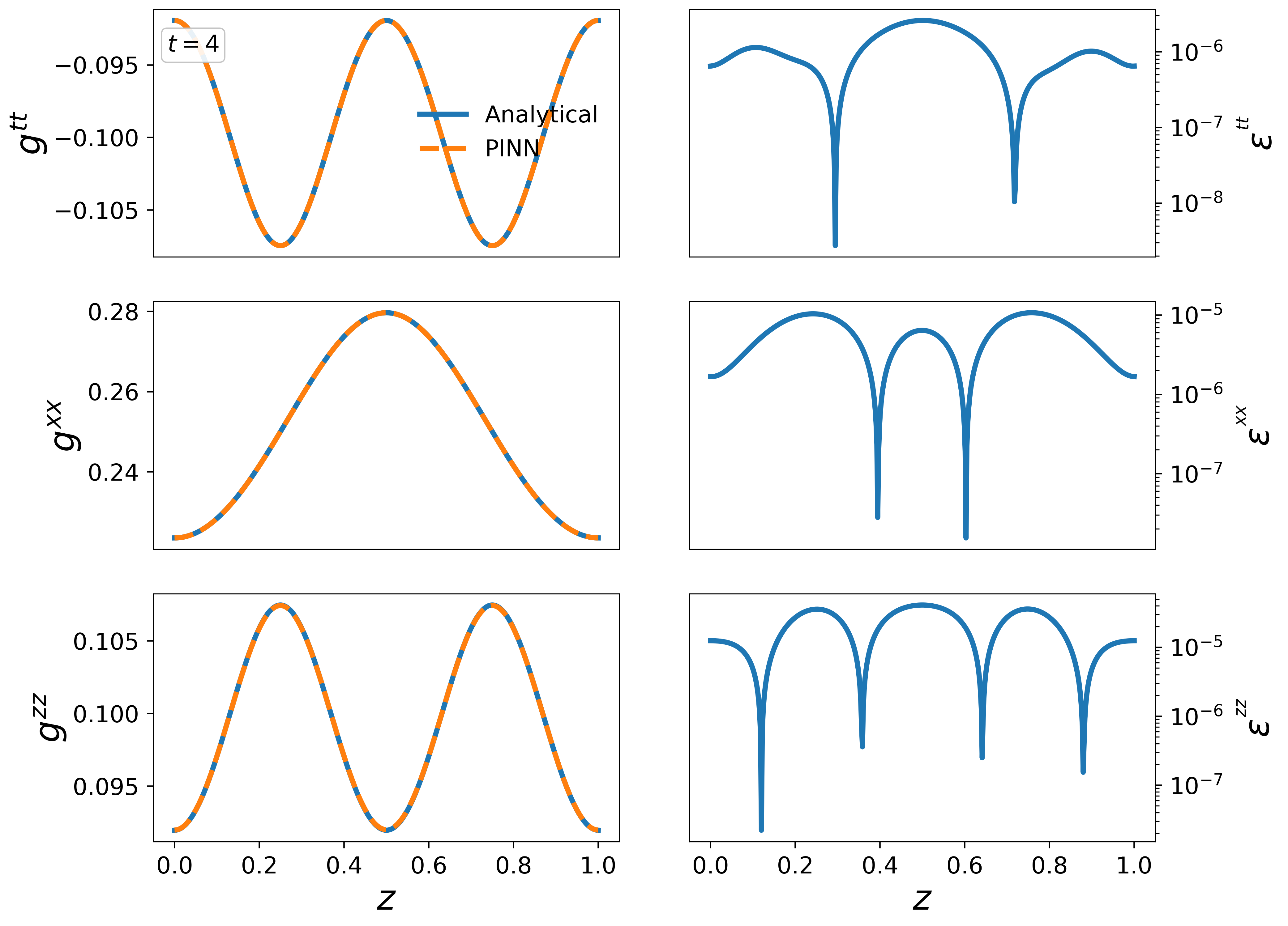}
    \caption{ {\em Gowdy spacetime test}.
        Comparison between the exact solution and the PINN prediction at
        $t=4$ during the expanding evolution. The selected metric components probe the oscillatory dependence generated by
        $P(t,z)$ and the exponential dependence on $\lambda(t,z)$. The
        corresponding absolute errors are of order $10^{-5}$ or smaller.
    }
    \label{fig:expansion}
\end{figure}
\end{widetext}

The expanding solution can be evolved reliably up to approximately $t=4$. Figure~\ref{fig:expansion} compares the PINN prediction with the exact solution at this time. The network accurately reproduces the spatial dependence of all three components, including the oscillatory structure associated with $P$ and the stronger variation generated by $\lambda$. The absolute errors remain of order $10^{-5}$ or smaller for all three components shown.

Beyond $t\approx4$, the optimization in the subsequent time windows fails to converge, and the PINN can no longer provide an accurate continuation toward the target time $t=8$. This breakdown, which has also been reported for finite-difference evolutions in the original benchmark~\cite{mexico}, suggests that the present training strategy has increasing difficulty in maintaining accuracy as the metric components develop increasingly different magnitudes and growth rates.

\onecolumngrid
\twocolumngrid

%%%%%%%%%%%%%%%%%%%%%%%%%%%%%%%%%%%%%%%%%%%%%%
\subsubsection{Collapsing Gowdy evolution}
\label{sec:gowdy_collapse}

For the collapsing Gowdy test, direct evolution toward $t=0$ would reach the cosmological singularity at a finite value of the time coordinate. We therefore adopt the harmonic slicing prescribed in the standard benchmark~\cite{mexico} and introduce a new time coordinate $\tau$ through
\begin{equation}
(t,x^i)
\longrightarrow
(\tau,x^i),
\qquad
t=F(\tau)=k\,\mathrm{e}^{c\tau},
\label{eq:gowdy_time_transformation}
\end{equation}
where $c$ and $k$ are positive constants.
\onecolumngrid
\begin{widetext}

\begin{figure}[H]
    \centering
    \includegraphics[width=0.7\linewidth]{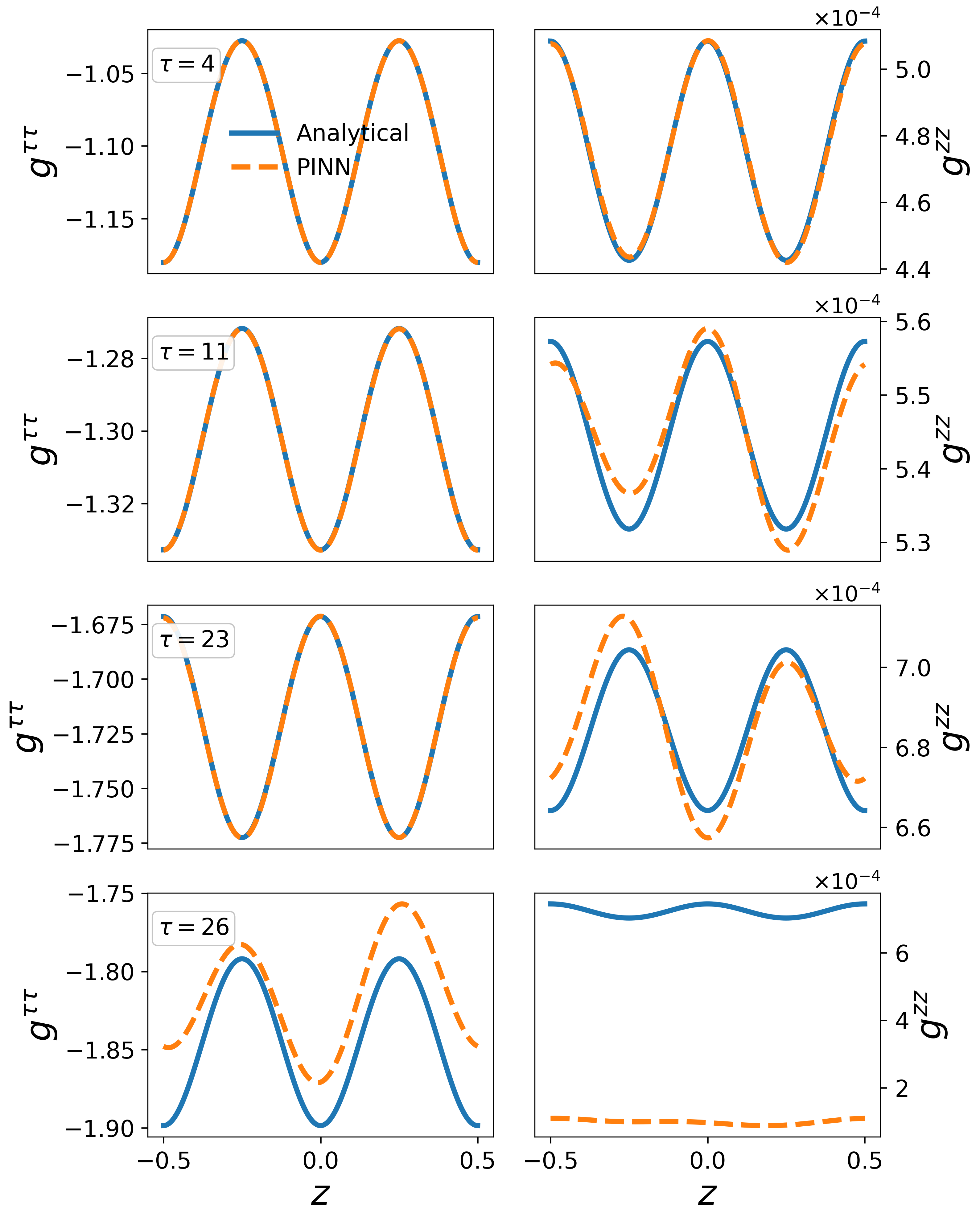}
    \caption{{\em Gowdy spacetime test}. Comparison between the exact collapsing solution and the PINN predictions at four representative times. The PINN accurately reproduces both components at early times. The first zero crossing of the oscillatory solution is followed by a persistent error in the smaller component, while deviations in the larger components become visible after the second zero crossing, leading to the breakdown observed at $\tau=26$.
    }
    \label{fig:gowdy_collapse_evolution}
\end{figure}
\end{widetext}
\twocolumngrid
Under this transformation, the singularity at $t=0$ is mapped to $\tau\rightarrow-\infty$, allowing the evolution to approach the singularity asymptotically.

In the transformed coordinates, we impose the corresponding gauge conditions directly on the inverse metric,
\begin{eqnarray}
g^{\tau i} &=& 0,
\\
g^{\tau\tau} &=&
-\frac{1}{c^2 k^{3/2}}
\exp\left(
-\frac{3c\tau}{2}
-\frac{\lambda(F(\tau),z)}{2}
\right).
\nonumber
\label{eq}
\end{eqnarray}
The constants $c$ and $k$ are chosen according to the standard prescription such that the initial slice satisfies the desired normalization of the time coordinate. Further details of the coordinate transformation and gauge choice are given in Ref.~\cite{mexico}.

A central difficulty in this test is the large disparity in magnitude among the inverse metric components. In particular, $g^{zz}$ becomes several orders of magnitude smaller than $g^{\tau\tau}$ and $g^{xx}$ during part of the evolution. Consequently, errors in $g^{zz}$ contribute relatively little to the total loss, making this component more difficult to resolve accurately.

Figure~\ref{fig:gowdy_collapse_evolution} compares the PINN predictions with the exact solution at four representative times. At $\tau=4$, the predictions are in close agreement with the exact solution for all three components, reproducing both their amplitudes and spatial dependence, including the much smaller $g^{zz}$ component.

The first significant deviation appears around $\tau\simeq12$, when the spatially varying contribution to $\lambda$, proportional to $J_0(2\pi T)J_1(2\pi T)$, vanishes at a zero of $J_1$ and subsequently changes sign. This change is most directly reflected in the metric components that depend on $\lambda$. The PINN continues to reproduce the main features of $g^{\tau\tau}$ and $g^{xx}$, but develops a noticeable amplitude error in the much smaller $g^{zz}$ component. This error persists at later times.

At $\tau=23$, $g^{\tau\tau}$ and $g^{xx}$ remain in good agreement with the exact solution, whereas $g^{zz}$ retains the correct qualitative spatial dependence but exhibits an incorrect amplitude. A further change occurs around $\tau\simeq24$, at the subsequent zero of $J_0$. At this point, $P$ passes through zero and changes sign, while the spatially varying contribution to $\lambda$ also vanishes because it contains the product $J_0J_1$. By $\tau=26$, the accumulated error in $g^{zz}$ has increased substantially, and noticeable deviations have also emerged in $g^{\tau\tau}$ and, to a lesser extent, in $g^{xx}$. The PINN therefore no longer provides an accurate representation of the exact solution. This suggests that the error initially concentrated in the smallest metric component accumulates across successive time windows and eventually propagates to the remaining components.
\begin{figure}[H]
    \centering
    \includegraphics[width=1.0\linewidth]{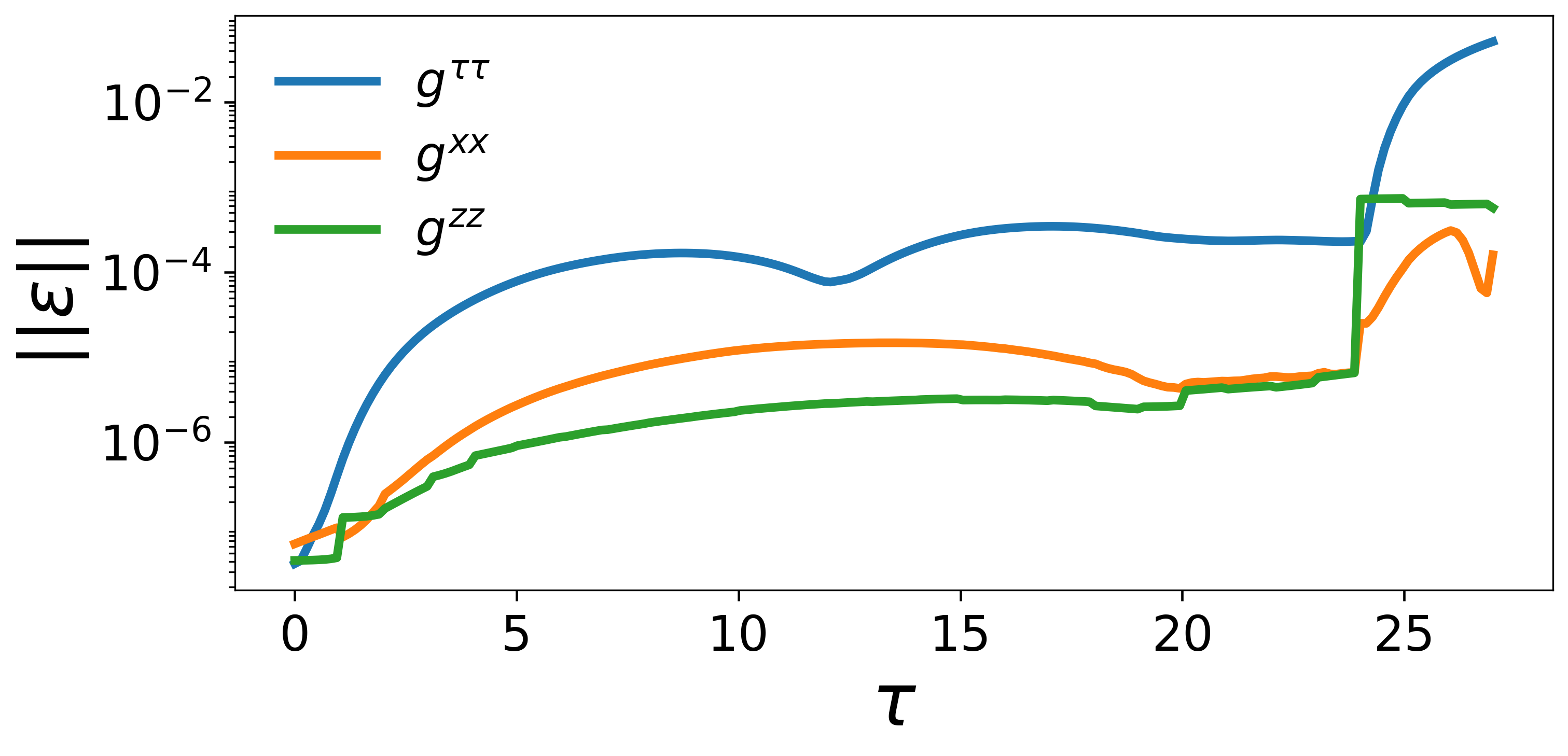}
    \caption{ {\em Gowdy spacetime test}.
    Evolution of the spatially averaged mean absolute error for the inverse metric components during the collapsing Gowdy test. The errors remain controlled up to approximately $\tau=24$. Around the second zero crossing of the oscillatory solution, the error in $g^{zz}$, whose characteristic magnitude is substantially smaller than those of the other components, increases abruptly. This is followed by a rapid growth of the errors in $g^{\tau\tau}$ and $g^{xx}$, consistent with the breakdown observed in Fig.~\ref{fig:gowdy_collapse_evolution}    }
    \label{fig:gowdy_collapse_errors}
\end{figure}

Figure~\ref{fig:gowdy_collapse_errors} quantifies this behavior through the spatially averaged absolute errors of the selected inverse metric components. For most of the evolution, the errors remain relatively small and vary gradually. Up to approximately $\tau=24$, the mean absolute errors in $g^{xx}$ and $g^{zz}$ remain $\lesssim 10^{-5}$, while the error in $g^{\tau\tau}$ stays below a few times $10^{-4}$.

Around the second zero of $J_0$, the error in $g^{zz}$ increases abruptly by more than two orders of magnitude. The errors in $g^{\tau\tau}$ and $g^{xx}$ then grow rapidly, with the error in $g^{\tau\tau}$ reaching values of order $10^{-2}$ near the end of the evolution. This sequence suggests that the loss of accuracy first becomes significant in $g^{zz}$ and subsequently propagates to the other metric components as the evolution proceeds.

The collapsing Gowdy test shows that the PINN can reproduce the nonlinear evolution over moderate times, including the first zero crossing of the oscillatory metric profiles. It also highlights the limitations associated with the large differences in magnitude among the metric components. An error that initially affects a small-amplitude component can persist across successive time windows and eventually propagate to the other components. Improving the long-term evolution may therefore require strategies that better balance the different components of the solution, such as component-wise normalization or adaptive loss weighting~\cite{wang2021gradient}. Network architectures better suited to representing high-frequency solutions may also be beneficial~\cite{wang2021fourier}. 

%%%%%%%%%%%%%%%%%%%%%%%%%%%%%%%%%%%%%%
\section{Three-Dimensional Matter-Coupled Test: Solitonic Boson Star}\label{sec:boson_star}
%%%%%%%%%%%%%%%%%%%%%%%%%%%%%%%%%%%%%%

Solitonic boson stars provide a natural extension of the validation of the PINN framework to non-vacuum spacetimes. They are regular, self-gravitating, and spatially localized solutions of the Einstein--Klein--Gordon system, providing a strong-field test in which the scalar field and spacetime geometry must be evolved simultaneously.

They also allow us to study nonlinear matter--gravity dynamics without  the additional complexities required for other compact objects: black hole simulations require a treatment of the horizon and singular interior, for example through excision~\cite{alcubierre2001excision} or puncture~\cite{campanelli2006evolutions}, while neutron star simulations require relativistic hydrodynamics and an equation of state~\cite{baiotti2017review}. By contrast, boson stars provide a comparatively simple matter model through a minimally coupled complex scalar field, making them a useful intermediate step toward more realistic matter--gravity systems.

As discussed previously, we consider a complex scalar field with the solitonic potential given in Eq.~\eqref{eq:solitonic_potential}. This potential admits highly compact boson-star configurations~\cite{revBS,solbs}, providing a suitable test of the PINN framework in the strong-gravity  regime, while avoiding discontinuities in the fields or singularities in the spacetime.

To complement the PINN results with an independent numerical reference, we have also evolved the same boson-star configuration using our finite-difference code, which uses the CCZ4 formulation of the Einstein equations~\cite{Alic:2011gg,solbs}. The simulations employ fourth-order Runge--Kutta time integration and a fourth-order finite-difference spatial discretization, with a CFL factor $\Delta t = 0.4\Delta x$. The highest resolution within the star is $\Delta x = 0.25$, corresponding to approximately 100 grid points across the stellar diameter. The computational domain extends to $L=\pm160$ in each spatial direction and uses five levels of fixed mesh refinement.

%%%%%%%%%%%%%%%%%%%%%%%%
\subsection{Initial and Boundary conditions}
%%%%%%%%%%%%%%%%%%%%%%%%%

%%%%%%%%%%%%%%%%%%%%%%%%%%%%%%%
\subsubsection{Initial data}
%%%%%%%%%%%%%%%%%%%%%%%%%%%%%%%

Equilibrium configurations of a self-gravitating complex scalar field can be obtained by assuming a static, spherically symmetric spacetime and a harmonic time dependence for the scalar field,
\begin{equation}
\Phi(t,r)=\varphi(r)e^{i\omega t},
\label{eq:boson_star_ansatz}
\end{equation}
where $\varphi(r)$ is a real radial profile and $\omega$ is the constant angular frequency. Despite its explicit time dependence, the scalar field has a stationary energy--momentum tensor, allowing the spacetime geometry to remain static.

We construct our initial data from one of these spherically symmetric equilibrium configurations. In isotropic Cartesian coordinates, the corresponding line element is given by
\begin{equation}
\begin{aligned}
ds^2
={}&
-\alpha^2(r)dt^2
+
\psi^4(r)
\left(
dx^2+dy^2+dz^2
\right),
\end{aligned}
\label{eq:boson_star_metric}
\end{equation}
where $\alpha(r)$ is the lapse and $\psi(r)$ is the conformal factor.
With this ansatz, the EKG equations reduce to a coupled system of ordinary differential equations for $\phi(r)$, $\alpha(r)$, and $\psi(r)$, with $\omega$ determined as an eigenvalue of the resulting boundary-value problem. Solving this equilibrium problem \cite{revBS,ff} is not part of the PINN calculation considered here. Instead, we use a numerical equilibrium solution obtained in Ref.~\cite{solbs}.

Although the equilibrium configuration is constructed assuming spherical symmetry, no symmetry reduction is imposed during the PINN evolution, which is performed in Cartesian coordinates. The PINN takes the full spacetime coordinates $(t,x,y,z)$ as inputs, and all derivatives entering the EKG residuals are computed by automatic differentiation with respect to these coordinates. Neither the network architecture nor the evolution equations make any explicit assumption of spherical symmetry. The PINN must therefore evolve the scalar field and spacetime geometry while preserving the localized, approximately spherical structure of the initial configuration.

The self-interaction potential is specified by the boson mass, $m_b=1$ and the solitonic parameter, $\sigma_0=0.05$. We choose an equilibrium configuration with a central field amplitude corresponding to a stellar compactness of $C=0.12$, following the convention adopted in Ref.~\cite{solbs}, and an eigenfrequency of $\omega=0.2673$. The physical units can be restored by rescaling with the boson mass. Here, setting $m_b=1$ fixes the characteristic scales of the problem and is also convenient numerically, as it keeps the boson-mass contribution to the Klein--Gordon equation of order unity and avoids introducing an additional scale disparity in the optimization.

In these units, the star has a characteristic radius of approximately $R_\star\approx12$, while the spherical computational domain extends to $R_{\rm max}=60$. Uniform volumetric sampling would therefore place only a small fraction of the collocation points inside the star, where the scalar field and the strongest curvature gradients are concentrated. We therefore increase the sampling density around the star using a Gaussian distribution centered at the origin, with  $\sigma=12$ . This provides better coverage of the dynamically relevant region while keeping the total number of collocation points manageable.

%%%%%%%%%%%%%%%%%%%%%%%%%%%%%%%
\subsubsection{Boundary conditions}
%%%%%%%%%%%%%%%%%%%%%%%%%%%%%%%

Since the evolution is performed in three spatial dimensions using Cartesian coordinates, boundary conditions must be imposed on the outer boundary of the computational domain. We choose this boundary to lie at a spherical surface, $r=R_{max}$, centered on the star and located in the weak-field region, where the scalar field is strongly suppressed and the spacetime approaches Minkowski space. We therefore impose approximate radiative Sommerfeld conditions~\cite{alcubierre} on both the scalar field and the metric.

For the real and imaginary components of the scalar field, we impose
\begin{equation}
\partial_t\Phi
+ \partial_r \Phi
+ \frac{\Phi}{r} = 0,
\label{eq:BC_phi}
\end{equation}
where the radial derivative $\partial_r=(x^i/r)\partial_i$ is expressed in Cartesian coordinates.

For the metric components, an equivalent radiative condition is imposed on their deviation from the Minkowski metric, $\delta g_{ab}=g_{ab}-\eta_{ab}$,
where $\eta_{ab}=\operatorname{diag}(-1,1,1,1)$. These conditions are imposed softly through the boundary contribution to the loss function and are intended to reduce spurious reflections of outgoing perturbations from the finite outer boundary.

%%%%%%%%%%%%%%%%%%%%%%%%
\subsection{Results of the time evolution}
%%%%%%%%%%%%%%%%%%%%%%%%%

The evolution of the boson star is performed using the sequential time-window decomposition, combined with the time-dependent exponential sampling law and the causal weighting scheme described in Section~\ref{subsec:multiwindow}. Since the boson star represents an equilibrium configuration, this provides a particularly clean test of the ability of the PINN to maintain the spatial structure of the solution over long evolution times.

\begin{widetext}
    
\begin{figure}[H]
    \centering
    \includegraphics[width=1.0\linewidth]{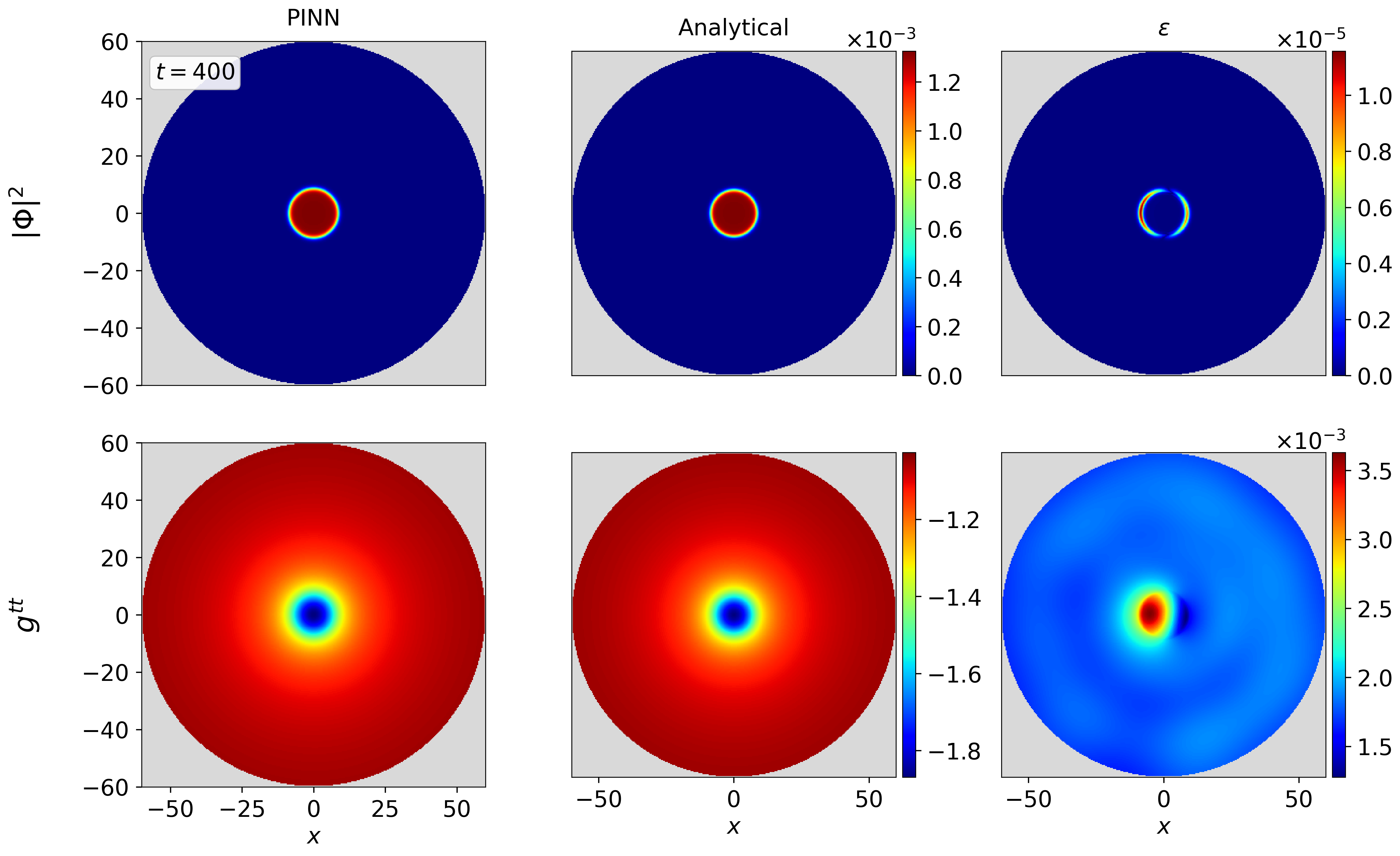}
    \caption{ {\em Solitonic boson star test}.
        Spatial structure of the scalar density $|\Phi|^2$ (top) and the inverse metric component $g^{tt}$ (bottom) on the equatorial plane $z=0$ at the final simulation time. The left and central columns show the PINN prediction and the equilibrium reference solution, respectively, while the right column shows the corresponding absolute errors. The PINN preserves the localized and approximately spherical matter distribution and reproduces the associated spacetime geometry throughout the evolution. The relative error in the scalar density remains below approximately $10^{-2}$, with its maximum near the surface of the star. The relative error in $g^{tt}$ is more spatially distributed and reaches values of order $10^{-3}$. 
    }
    \label{fig:bs_final_slice}
\end{figure}

\end{widetext}

Figure~\ref{fig:bs_final_slice} compares the PINN prediction with the reference solution on the equatorial plane $z=0$ at the final simulation time $t=400$, corresponding to approximately 17 cycles of the scalar field. The rows show the scalar density $|\Phi|^2$ and the inverse metric component $g^{tt}$, while the right column shows the corresponding absolute errors.
\onecolumngrid
\twocolumngrid

Although the real and imaginary parts of the scalar field undergo the harmonic phase rotation of Eq.~\eqref{eq}, the scalar density $|\Phi|^2$ remains stationary in the exact equilibrium solution. It therefore provides a direct diagnostic of how well the PINN preserves the localized matter distribution. The PINN captures both the approximately constant central core and the rapid decay toward the exterior, with the largest discrepancies occurring near the stellar surface, where the scalar profile has its strongest spatial gradients. Nevertheless, the absolute error remains at most $\mathcal{O}(10^{-5})$ across the computational domain.

The inverse metric component $g^{tt}$ provides a complementary diagnostic of the spacetime geometry. Its spatial profile extends beyond the matter distribution and approaches its asymptotic value away from the center. At the final time $t=400$, the PINN reproduces this global structure, with absolute errors of order $10^{-3}$ that are more spatially extended than those of the scalar field.
\begin{figure}[H]
    \centering
    \includegraphics[width=\columnwidth]
    {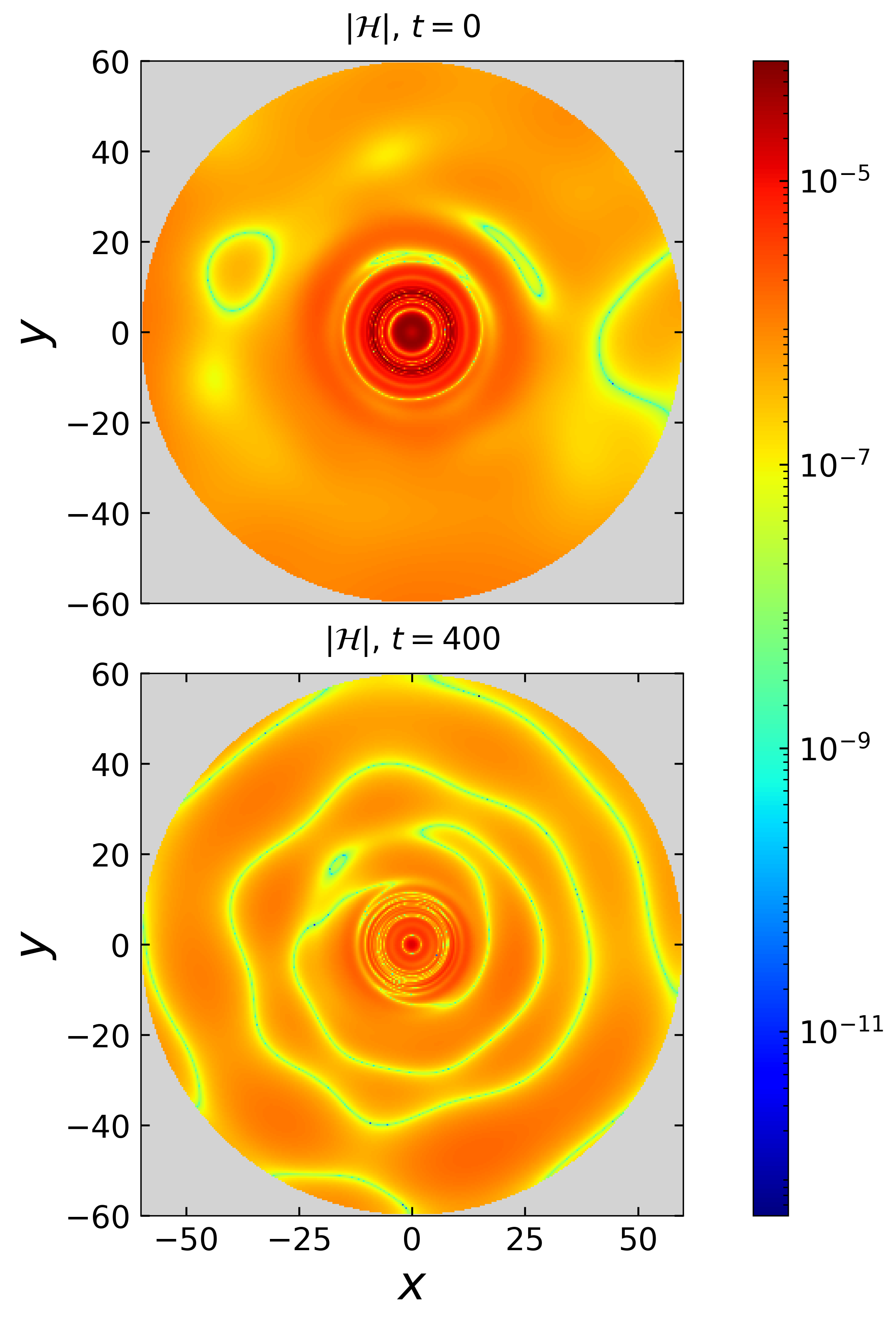}
    \vspace{-2em}
    \caption{ {\em Solitonic boson star test}.  
    Magnitude of the Hamiltonian-constraint residual, $|\mathcal{H}|$, on the equatorial plane $z=0$ at the initial and final times of the simulations. The residual is initially larger throughout the stellar interior, whereas at late times it becomes more broadly distributed across the domain.
    }
    \label{fig:hamiltonian_constraint}
\end{figure}

Another useful measure of how accurately the Einstein equations are satisfied is the Hamiltonian-constraint residual. This can be obtained by projecting the Einstein residual twice along the future-directed unit normal $n^a$ to the spatial hypersurfaces, namely
\begin{equation}
\mathcal{H}
=
\left[
R_{ab} -8\pi ( T_{ab}- \frac{1}{2}T g_{ab})
\right]
n^a n^b.
\label{eq:hamiltonian_residual}
\end{equation}
For an exact solution of the Einstein--Klein--Gordon system, this quantity vanishes identically, so its magnitude provides an independent measure of the accuracy of the gravitational sector.
Since the Hamiltonian and momentum constraints are coupled through their propagation, monitoring the Hamiltonian constraint already provides a useful indirect check of the consistency of the full constraint system during the evolution.

Figure~\ref{fig:hamiltonian_constraint} shows $|\mathcal{H}|$ on the equatorial plane $z=0$ at the initial and final times. At $t=0$, the residual is mainly concentrated within the star and around regions with strong matter and metric gradients. By $t=400$, the constraint violation is reduced in the central region and becomes more spatially distributed. No significant increase in the residual is observed toward the outer part of the domain, including near the boundary where the approximate radiative conditions are imposed.

\begin{figure}[H]
    \centering
    \includegraphics[width=\columnwidth]
    {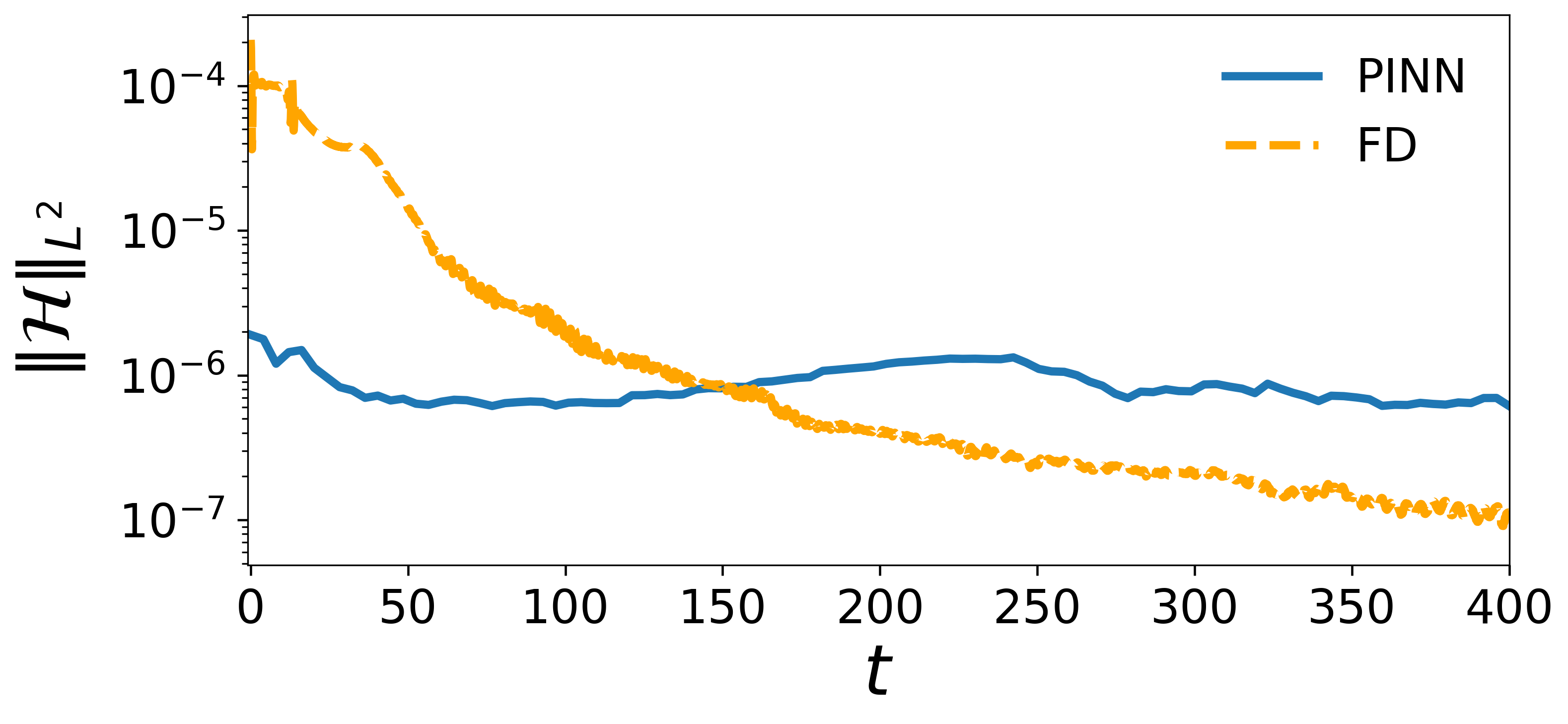}
    \caption{{\em Solitonic boson star test}.  
    $L_2$-norm of the Hamiltonian-constraint residual during the simulation. The finite-difference residual decreases during the evolution as the constraint violations propagate out of the domain. 
     }
 \label{fig:hamiltonian_constraint_L2norm}
\end{figure}

The decrease in the Hamiltonian-constraint residual, more clearly seen in the $L_2-$norm shown in Figure~\ref{fig:hamiltonian_constraint_L2norm}, can be understood from the soft enforcement of the initial numerical data and the inclusion of the constraint equations directly in the PDE loss. The initial configuration contains a small constraint violation inherited from its numerical construction and interpolation. Because the initial conditions are imposed softly, the optimization can introduce small deviations from the initial data while minimizing the EKG residuals. The resulting PINN solution may therefore be closer to the constraint-satisfying solution than the original numerical data.
As the evolution proceeds, the initially localized constraint violation may propagate and become distributed over a larger portion of the computational domain, thereby reducing its local amplitude while maintaining a similar global norm of the error. The finite-difference solution exhibits larger constraint violations initially but substantially smaller violations at late times, largely because it uses the CCZ4 formulation, which incorporates constraint propagation by construction.\footnote{The dynamical constraint-damping mechanisms that actively drive constraint violations toward zero are disabled in order to match as closely as possible the Einstein equations solved by the PINN.} 

The total Noether charge associated with the $U(1)$ symmetry of the complex scalar field provides a quantitative global diagnostic. For the exact equilibrium solution, this quantity is conserved throughout the evolution, making it a useful measure of whether the PINN preserves the global matter content of the boson star.

\begin{figure}[H]
    \centering
    \includegraphics[width=1.0\linewidth]{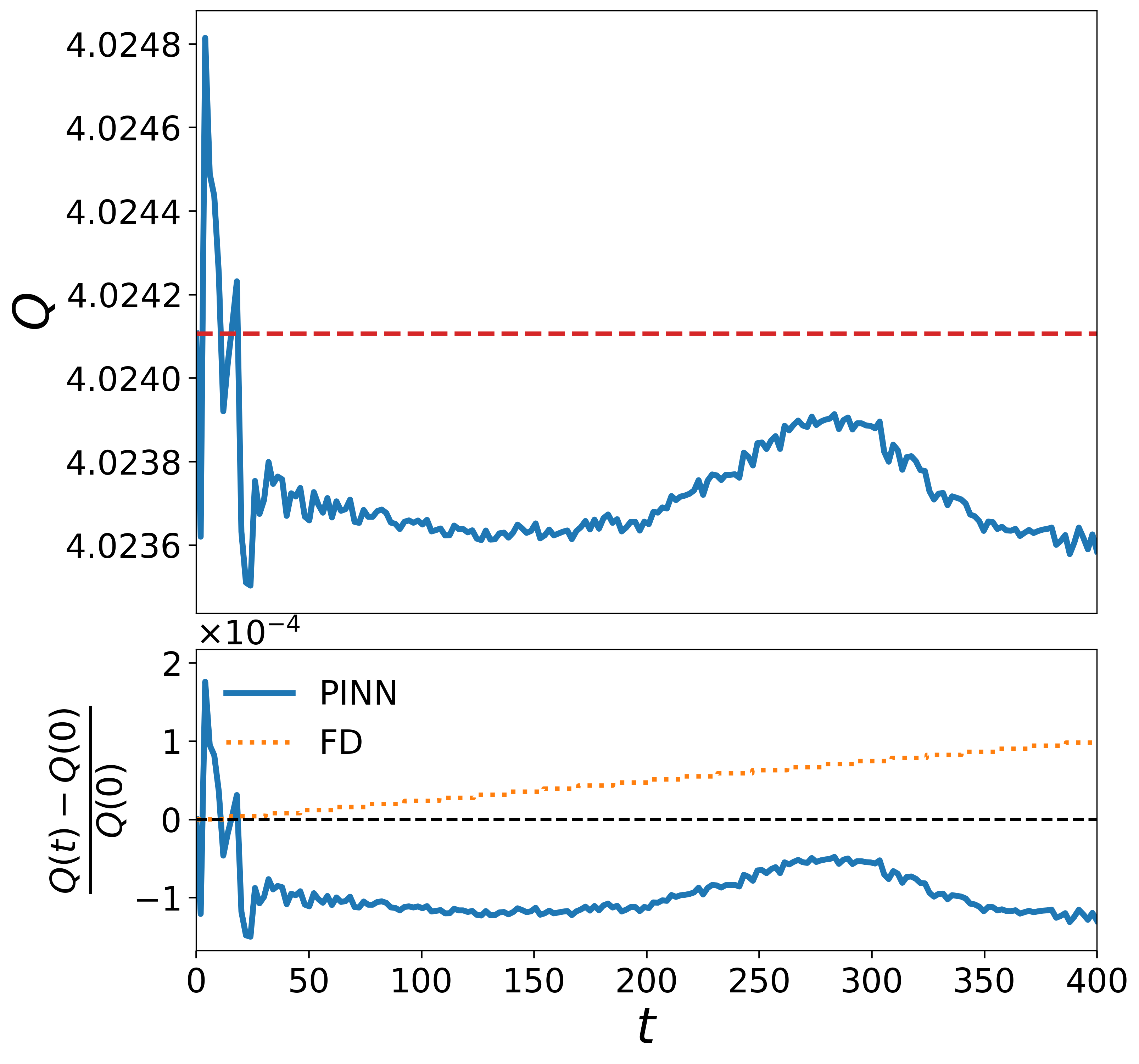}
    \caption{ {\em Solitonic boson star test}.  Evolution of the Noether charge and its relative variation as a function of time. The charge exhibits small initial fluctuations, after which its relative deviation remains approximately $\mathcal{O}(10^{-4})$ throughout the evolution, comparable to the corresponding variation in the finite-difference simulation.
    }    \label{fig:noether_evolution}
\end{figure}

Figure~\ref{fig:noether_evolution} shows the evolution of the Noether charge up to $t=400$. The upper panel compares the PINN value with the initial charge, while the lower panel shows the corresponding relative variation. The charge remains close to its initial value throughout the simulation, with the relative deviation remaining below approximately $10^{-4}$. For comparison, the corresponding deviation in the finite-difference simulation is approximately of the same order of magnitude.

A further diagnostic is provided by the scalar field at the center of the star. For the equilibrium configuration considered here, the complex scalar field follows the harmonic dependence of Eq.~\eqref{eq:boson_star_ansatz}. At the origin, it therefore takes the form $\phi(0)e^{i\omega t}$, so that its real and imaginary parts oscillate with constant amplitude and a phase difference of $\pi/2$.

\begin{figure}[H]
    \centering
    \includegraphics[width=1.0\linewidth]{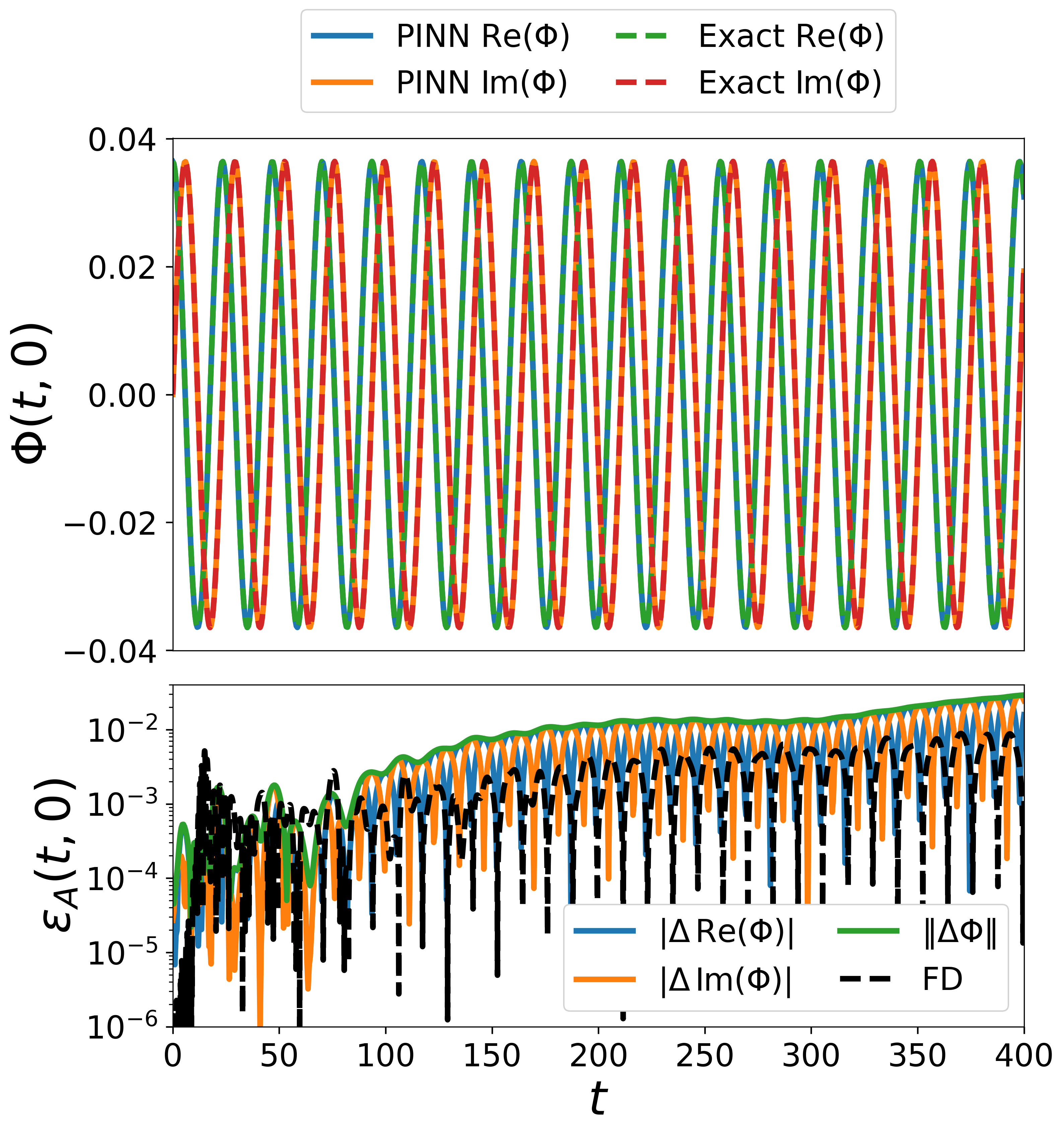}
   \caption{ {\em Solitonic boson star test}. (Top) Time evolution of the real and imaginary components of the scalar field at the center of the boson star, showing the expected harmonic behavior with frequency $\omega$. (Bottom) Relative errors, normalized by the maximum scalar-field amplitude, in $\mathrm{Re}(\Phi)$ and $\mathrm{Im}(\Phi)$, together with the magnitude of the complex scalar-field error obtained from the PINN and the finite-difference simulation.}
    \label{fig:corebs}
\end{figure}

Figure~\ref{fig:corebs} compares the PINN prediction with the reference solution for $\mathrm{Re}(\Phi)$ and $\mathrm{Im}(\Phi)$ at the origin over the full evolution. The upper panel shows that the PINN reproduces both the oscillation frequency and the relative phase, with no visible phase drift. The lower panel shows the corresponding absolute errors in the two components and in the complex field. The relative errors remain of order $10^{-3}$ throughout the evolution, with only a mild increase at late times. For these quantities, the error of the finite-difference simulation is comparable to that of the PINN.

Finally, we investigate the characteristic oscillation modes of the boson star. These modes can be excited by small deviations of the PINN solution from the exact equilibrium configuration, arising from finite training accuracy, sampling, and the numerical representation of the initial data. Their frequencies are of particular interest because they encode information about the internal structure and stability of the object, potentially providing observational signatures that could distinguish boson stars from other compact objects~\cite{QNM} and between different boson-star configurations~\cite{solbs}. Recovering the expected oscillation spectrum therefore provides an additional, nontrivial validation that the PINN captures the underlying dynamics of the EKG system.

\begin{figure}[H]
    \centering
    \includegraphics[width=1.0\linewidth]{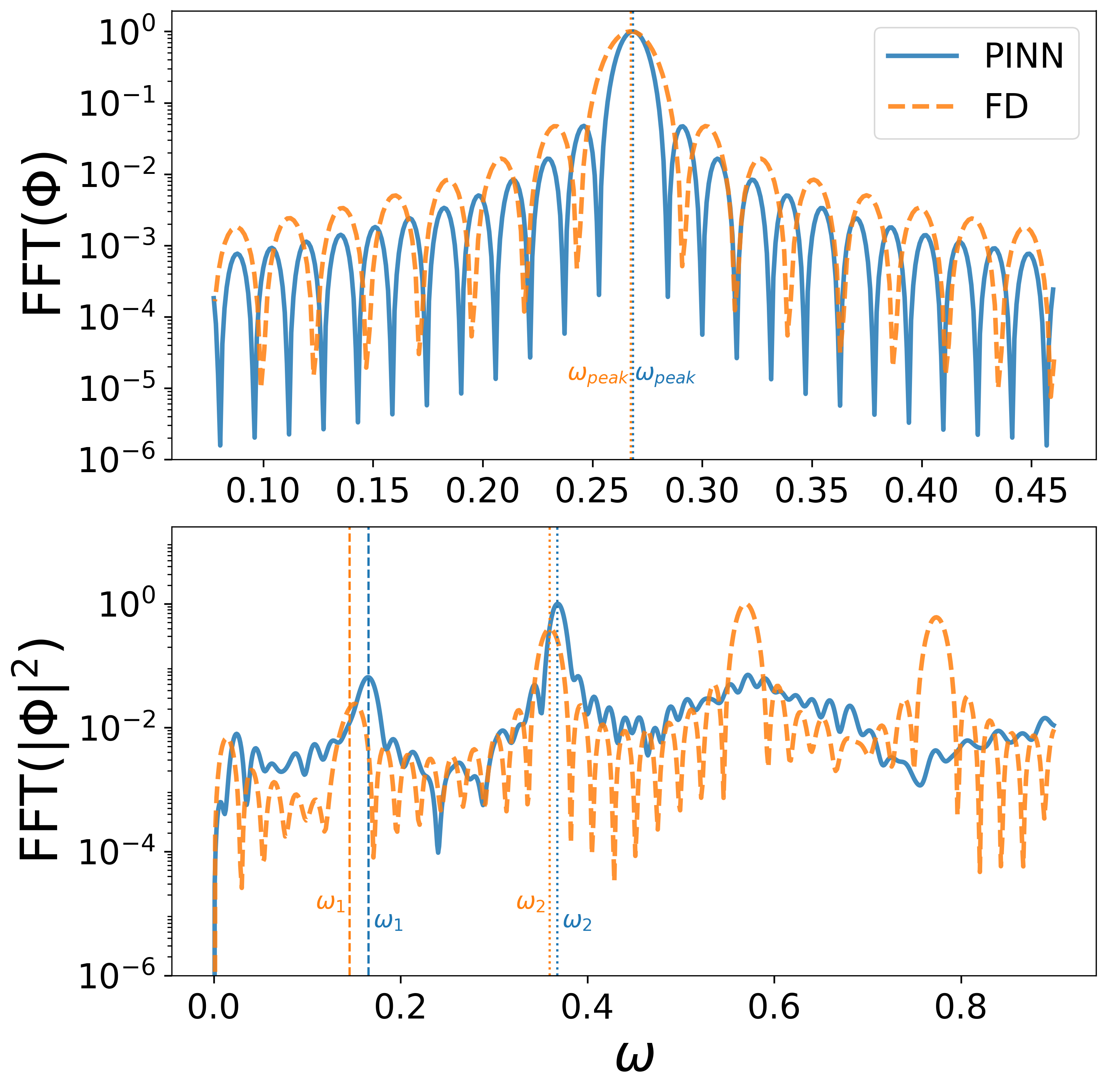}
    \caption{{\em Solitonic boson star test}. Comparison of the normalized power spectra obtained from the PINN and finite-difference evolutions. (Top) Power spectrum of the complex scalar field $\Phi(t,0)$, whose dominant peak corresponds to the harmonic phase frequency of the equilibrium configuration. (Bottom) Power spectrum of the variations in $|\Phi(t,0)|^2$, revealing the characteristic radial oscillation frequencies of the star. The reasonable agreement of these frequencies, listed in Table~\ref{tab:QNM_frequencies}, demonstrates that the PINN reproduces the oscillatory dynamics obtained with the reference finite-difference evolution. 
    }
    \label{fig:QNM}
\end{figure}

Figure~\ref{fig:QNM} compares the normalized power spectra obtained from the PINN and finite-difference evolutions~\cite{solbs}. The upper panel shows the spectrum of the complex scalar field $\Phi(t,0)$, which is dominated by the intrinsic harmonic dependence of the equilibrium solution at the frequency $\omega_{\rm peak}$. The lower panel shows the spectrum of the variations in $|\Phi(t,0)|^2$, which isolates the physical pulsations of the star, corresponding to its oscillation modes.The intrinsic harmonic frequency and the two lowest-frequency oscillation modes are listed in Table~\ref{tab:QNM_frequencies}, showing reasonably good agreement with the corresponding finite-difference results.

\begin{table}[H]
    \centering
    \caption{{\em Solitonic boson star test}. Characteristic frequencies obtained from the PINN and finite-difference evolutions of the boson star. The frequency $\omega_{\rm peak}$ denotes the harmonic frequency of the complex scalar field $\Phi$, with $\omega_{\rm peak}=0.2673$ in the initial data. The frequencies $(\omega_1,\omega_2)$ correspond to the two lowest radial oscillation modes, identified from the spectrum of the variations in $|\Phi(t,0)|^2$.}
    \label{tab:QNM_frequencies}
    \begin{tabular}{lcc}
        \hline
        Frequency & PINN & Finite Difference \\
        \hline
        $\omega_{\rm peak}$ & $0.2684$ & $0.2674$ \\
        $\omega_1$           & $0.1569$ & $0.1657$ \\
        $\omega_2$           & $0.3677$ & $0.3596$ \\
        \hline
    \end{tabular}
\end{table}

%%%%%%%%%%%%%%%%%%%%%%%%%%%%%%%%%%%%%%%%%%%%%%%%%
\section{Conclusions}
\label{sec:conclusions}
%%%%%%%%%%%%%%%%%%%%%%%%%%%%%%%%%%%%%%%%%%%%%%%%

In this work, we have investigated the use of PINNs to solve the Einstein--Klein--Gordon system. 
Our formulation enforces the covariant equations directly in terms of the four-dimensional metric and complex scalar field, without introducing auxiliary variables for  an order reduction of the system, as is typically done in standard $(3+1)$ formulations.  
The gauge conditions are incorporated directly into the system. The network is therefore trained using the ten Einstein equations, the two components (i.e., the real and imaginary parts) of the Klein--Gordon equation, and four equations specifying the gauge conditions. 
Our results show that accurate time-dependent solutions can be obtained by enforcing this strongly coupled system directly in its covariant differential form.

We first consider a sequence of vacuum benchmarks to assess the behavior of the method under increasingly complex conditions. The linear-wave and gauge-wave experiments show that the PINN can reproduce propagating solutions over many crossing times, while the Gowdy tests expose some of the current limitations of the approach. This is a particularly demanding benchmark even for traditional numerical methods, owing to the increasingly disparate characteristic scales that must be resolved simultaneously. For a single neural representation, this requires capturing features with substantially different spatial, temporal, and field-amplitude scales within the same model. These results indicate that handling strongly multiscale dynamics remains an important limitation of the current PINN-based evolution strategy.

The most stringent test presented in this work is the evolution of a solitonic boson star with the complete Einstein--Klein--Gordon system. Despite the compactness of the configuration and the nonlinear coupling between the scalar field and the geometry, the PINN preserves the main properties of the equilibrium solution throughout the evolution considered here. 
The scalar-field phase and amplitude remain within the reported numerical errors, while the spacetime geometry remains close to the reference solution.
The agreement is not restricted to local field values: the Noether charge, which measures the conserved matter content, varies by approximately \(10^{-4}\) over the complete evolution.

An independent evolution of the same configuration using a high-resolution finite-difference code provides a further quantitative comparison. The PINN reproduces the spatial structure and time dependence of the scalar field and metric, with errors of comparable magnitude to those obtained in the finite-difference evolution for several of the diagnostics considered. As a further test, we consider the oscillation spectrum generated by small deviations from the equilibrium configuration. These perturbations excite radial oscillation modes, whose characteristic frequencies can be identified from the power spectrum of the scalar density. 
The intrinsic field frequency agrees with the finite-difference results at the sub-percent level, while the two lowest radial-mode frequencies are recovered with lower accuracy. This shows that the network might capture not only the equilibrium configuration but also the characteristic dynamical response of the coupled matter--geometry system.

Several numerical choices were found to play an important role in the robustness of the method. 
Following the systematic optimizer comparisons of Ref. \cite{urban2025unveiling}, we employed the SSBroyden quasi-Newton method throughout this work. This optimization strategy enabled the equation residuals to be reduced to the accuracy required for the long-time relativistic evolutions presented here.
The treatment of the initial data was also important for long-time evolution: soft enforcement proved more robust than hard temporal constructions, for which the error tended to grow systematically with time. Sampling strategies that preserve the causal structure of the evolution, together with the corresponding causal weighting in the loss, were likewise beneficial.
Finally, appropriate rescaling of the coordinates and fields to dimensionless variables was important to avoid saturating the activation functions and thereby improve the conditioning of the optimization problem.

Beyond their ability to solve the equations, PINNs offer features that differ qualitatively from conventional numerical methods. One such feature is the possibility of representing a continuous family of solutions over a range of physical parameters with a single training procedure. This is illustrated by the parametric gauge-wave experiment, in which the wave amplitude is introduced as an additional network input. A single training therefore learns the dependence of the solution on the amplitude over the prescribed parameter interval, so that different members of the family can subsequently be obtained simply by changing the value of the input parameter.
In contrast, a conventional finite-difference or spectral code would generally require a separate simulation for each parameter value. Parametric PINNs may therefore be particularly useful, and potentially competitive, for applications requiring the exploration of broad regions of parameter space.

A second advantage lies in the combination of automatic differentiation with the flexibility to construct the loss function directly from the governing equations. We discuss these two aspects separately below. 
Automatic differentiation evaluates derivatives of the neural representation without introducing finite-difference truncation errors or requiring explicit differentiation stencils or auxiliary variables to reduce higher-order equations to first-order form.
For example, theories involving higher-order spatial or temporal derivatives could, in principle, be incorporated directly into the PINN formulation. This does not remove the numerical difficulties associated with higher-order or stiff equations, but it can substantially reduce the algebraic and implementation effort required to formulate them.
The loss function provides a direct formal mechanism for combining equations and constraints of different types, although their relative scaling and optimization remain nontrivial. 
PINNs may therefore be useful as a rapid prototyping tool for exploring new models and formulations before developing dedicated high-accuracy numerical-relativity codes.

Finally, a further practical advantage comes from the stochastic sampling of collocation points. The equations do not need to be evaluated simultaneously on a dense multidimensional grid. Instead, relatively small batches of collocation points can be repeatedly resampled during training, allowing a large fraction of the continuous domain to be explored without storing a correspondingly large grid. This becomes particularly relevant as the dimensionality of the problem increases. In our three-dimensional boson-star simulation, both memory usage and computational cost increased by approximately an order of magnitude relative to the corresponding one-dimensional tests. Although higher-dimensional problems remain computationally demanding, the PINN formulation avoids the severe memory requirements that would arise from representing the full spacetime domain on a dense Cartesian grid.

The present approach nevertheless has some significant limitations. Most notably, the computational cost of training remains substantially higher than that of mature finite-difference or spectral methods for problems where efficient numerical codes are already available. PINNs are a comparatively recent approach, whereas the numerical methods underlying current numerical-relativity codes have benefited from decades of theoretical development, algorithmic refinement, and optimization for modern computing architectures. Moreover, the implementation developed here was designed primarily as a proof of concept rather than as a fully optimized solver, leaving considerable room for improvement in computational efficiency.
The results presented should therefore be viewed primarily as a demonstration of the feasibility of applying PINNs to these problems, rather than as evidence of numerical competitiveness with established numerical-relativity methods.

Despite these limitations, PINNs provide a flexible framework for exploring problems in numerical relativity. Their ability to work directly with higher-order equations of mixed type, their mesh-free formulation, and the possibility of representing parameterized families of solutions with a single network open possibilities that are less natural in conventional approaches. 
Future work should focus on improving the representation of multiscale solutions, transferring information more robustly between temporal windows, and reducing the computational cost of training.
Progress in these directions will be important for applying PINNs to increasingly complex problems in numerical relativity.

%%%%%%%%%%%%%%%%%%%%%%%%%%%%%%%%%%%%%%%%%%%%%%%%%%%%%%%%%%%%%%%%%%
%%%%%%%%%%%%%%%%%%%%%%%%%%%%%%%%%%%%%%%%%%%%%%%%%%%%%%%%%%%%%%%%%%
%%%%%%%%%%%%%%%%%%%%%%%%%%%%%%%%%%%%%%%%%%%%%%%%%%%%%%%%%%%%%%%%%%
%%%%%%%%%%%%%%%%%%%%%%%%%%%%%%%%%%%%%%%%%%%%%%%%%%%%%%%%%%%%%%%%%%

\section{Acknowledgments}
%%%%%%%%%%%%%%%%%%%%%%%%%%%%%%%%%%%%%%%%%%%%%%%%%%%%%%%%%%%%%%%%%%

This work was supported by the Project No. PID2022-138963NB-I00 and PID2025-168215NB-I00, funded by the Spanish Ministry of Science,
Innovation and Universities (MCIN/AEI/10.13039/
501100011033). J.A.C. acknowledges support from a
predoctoral fellowship (Formación de Personal
Investigador (FPI)) associated with this project (reference
No. PREP2022-000480). 
J.A.P acknowledges financial support from the Generalitat Valenciana through PROMETEO PROJECT
CIPROM/2022/13 and from MCIN/AEI through Project PID2025-171322NB-C21. 
The authors thankfully acknowledge the computer
resources at MareNostrum and the technical support
provided by Barcelona Supercomputing Center (RES-AECT-2025-2-0024 and RES-AECT-2026-1-0002). The authors acknowledge the computational resources and assistance provided by the TalaIA high-performance computing cluster at the University of the Balearic Islands (UIB) / Balearic Islands Center for Supercomputing and Artificial Intelligence (BSAI). The authors acknowledge the computational resources and assistance provided by UNC Supercómputo (CCAD), Universidad Nacional de Córdoba, Argentina.

\appendix
\section{Optimization}
\label{app:optimization}

% Frase inicial para resolver la discusión, para más información, ir al origen.
The purpose of this appendix is to summarize the optimization algorithm used in the present calculations, rather than to repeat the systematic optimizer comparison performed in 
\cite{urban2025unveiling}.
That study compared Adam, BFGS, and self-scaled Broyden strategies across several systems of differential equations and found that the latter achieved substantially smaller residuals and solution errors. We therefore adopt the same SSBroyden prescription here, which we briefly review next.

To solve problem~\eqref{eq:nn_eq}, the trainable parameters are updated iteratively according to a prescribed mathematical rule that generates a sequence of progressively better approximations to the solution. This iterative procedure defines the optimization algorithm, also known as the \emph{optimizer}. In this work, we consider a \emph{line-search} algorithm, in which the parameters are updated according to \cite{NocedalWright2006}:
\begin{equation}
\theta_{k+1} = \theta_k + \alpha_k p_k,
\end{equation}
where $p_k$ is a vector with the same dimension as $\theta_k$, and $\alpha_k$ is a scalar \emph{step length}.

To ensure that the loss function decreases in every iteration, we require $p_k$ to be a descent direction, i.e., the loss function must decrease locally along $p_k$. To this end, we define
\begin{equation}\label{eq:pk}
p_k = -H_k \nabla \mathcal{L}(\theta_k),
\end{equation}
where $H_k$ is a symmetric positive-definite matrix, so that $p_k^T \nabla \mathcal{L}(\theta_k) <0$. The simplest choice that guarantees a descent direction is $H_k = I$, where $I$ denotes the identity matrix. The resulting algorithm defines the well-known gradient descent method \cite{NocedalWright2006}. Other choices that have become popular in deep learning include Momentum gradient descent \cite{polyak1964speeding}, RMSprop \cite{RMSprop}, and Adam \cite{adam}, which can be interpreted as variants of gradient descent that implicitly modify the search direction by exploiting information from previous iterations, such as accumulated gradients or adaptive scaling.

While the previous methods are attractive due to their relatively low computational cost, it is becoming increasingly clear that optimization algorithms based on Newton's method, in which $H_k$ is chosen as the exact inverse of the Hessian matrix of the loss function, are considerably more effective for training PINNs \cite{urban2025unveiling,kiyani2025optimizer,JNINI2026106955}. An attractive example is provided by \emph{quasi-Newton} methods, where $H_k$ is an approximation to the inverse Hessian matrix that is constantly updated, together with the trainable parameters, so as to progressively improve this approximation using gradient information \cite{NocedalWright2006}. 

A general class of quasi-Newton updates can be expressed in terms of the \emph{self-scaled Broyden} formula \cite{urban2025unveiling}. If we define the auxiliary variables  
\begin{align}
    s_k &= \theta_{k+1} - \theta_k, \label{eq:sk_chap4} \\
    y_k &= \nabla \mathcal{L}(\theta_{k + 1}) - \nabla \mathcal{L}({\theta}_k), \\
    v_k &= \sqrt{{y}_k^T H_k y_k}\left[ \frac{s_k}{y_k^T s_k} - \frac{H_k y_k}{y_k H_k y_k}\right],
\end{align}
the next approximation of the inverse Hessian matrix at each iteration can be calculated by
\begin{equation}\label{eq:self_scaled_broyden}
    H_{k+1} = \frac{1}{\tau_k} \left[H_k - \frac{H_k y_k y_k^T H_k}{y_k^T H_k y_k} + \phi_k v_k v_k^T \right] + \frac{s_k s_k^T}{y_k^T s_k}, 
\end{equation}
where $\tau_k, \phi_k$ are the scaling and updating parameters, respectively, which in general change between iterations. In this work, we consider the parameter choices employed by the \emph{SSBroyden} optimizer proposed in \cite{urban2025unveiling}, which has been shown to be highly effective across a wide range of physical problems. 

Finally, we must specify the procedure used to determine the step length $\alpha_k$ at each iteration. In this work, we employ a backtracking line-search strategy \cite{NocedalWright2006}. Given the descent direction defined by Eq.~\eqref{eq:pk}, this procedure generates a sequence of candidate step lengths until one satisfying the \emph{sufficient decrease} (Armijo) condition \cite{Armijo1966MinimizationOF} is found:
\begin{equation}
\mathcal{L}(\theta_k + \alpha_k p_k) \leq \mathcal{L}(\theta_k) + c\alpha_k p_k^T\nabla \mathcal{L}(\theta_k),
\end{equation}
where $c \in (0,1)$ is a prescribed constant. Starting from an initial trial step length, successive candidates are obtained through quadratic and cubic interpolation of the objective function, while restricting the reduction of the step length to a prescribed interval (see \cite{dennis1996numerical} for details). The procedure terminates as soon as the Armijo condition is satisfied.
\section{Hard and soft enforcement of the initial data}
\label{app:hard_soft}
To determine how the initial conditions should be incorporated into the PINN, we compared hard and soft enforcement using the one-dimensional wave equation,
\begin{equation}
    \partial_t^2u -c^2\partial_x^2u=0
    \label{eq:hard_soft_wave}
\end{equation}
with initial data 
\begin{equation} u(t_0,x)=u_0(x), \qquad \partial_tu(t_0,x)=v_0(x). 
\end{equation}
For hard enforcement, the PINN approximation was constructed as
\begin{equation}
\hat{u}^{\mathrm{hard}}_\theta(t,x)
= u_0(x) +(t-t_0)v_0(x) +(t-t_0)^2 \mathcal{N}_\theta(t,x),
\label{eq:hard_temporal_ansatz}
\end{equation}
where \(\mathcal{N}_\theta(t,x)\) denotes the raw output of the neural network, with \(\theta\) representing its trainable parameters. This construction satisfies both initial conditions identically for all values of \(\theta\). In the soft formulation, the network output was used directly as the PINN approximation,
\begin{equation}
\hat{u}^{\mathrm{soft}}_\theta(t,x) = \mathcal{N}_\theta(t,x),
\end{equation}
and  deviations from the prescribed initial data were included as additional contributions to the loss function.

\begin{figure}[H] 
    \centering \includegraphics[width=1.0\columnwidth]{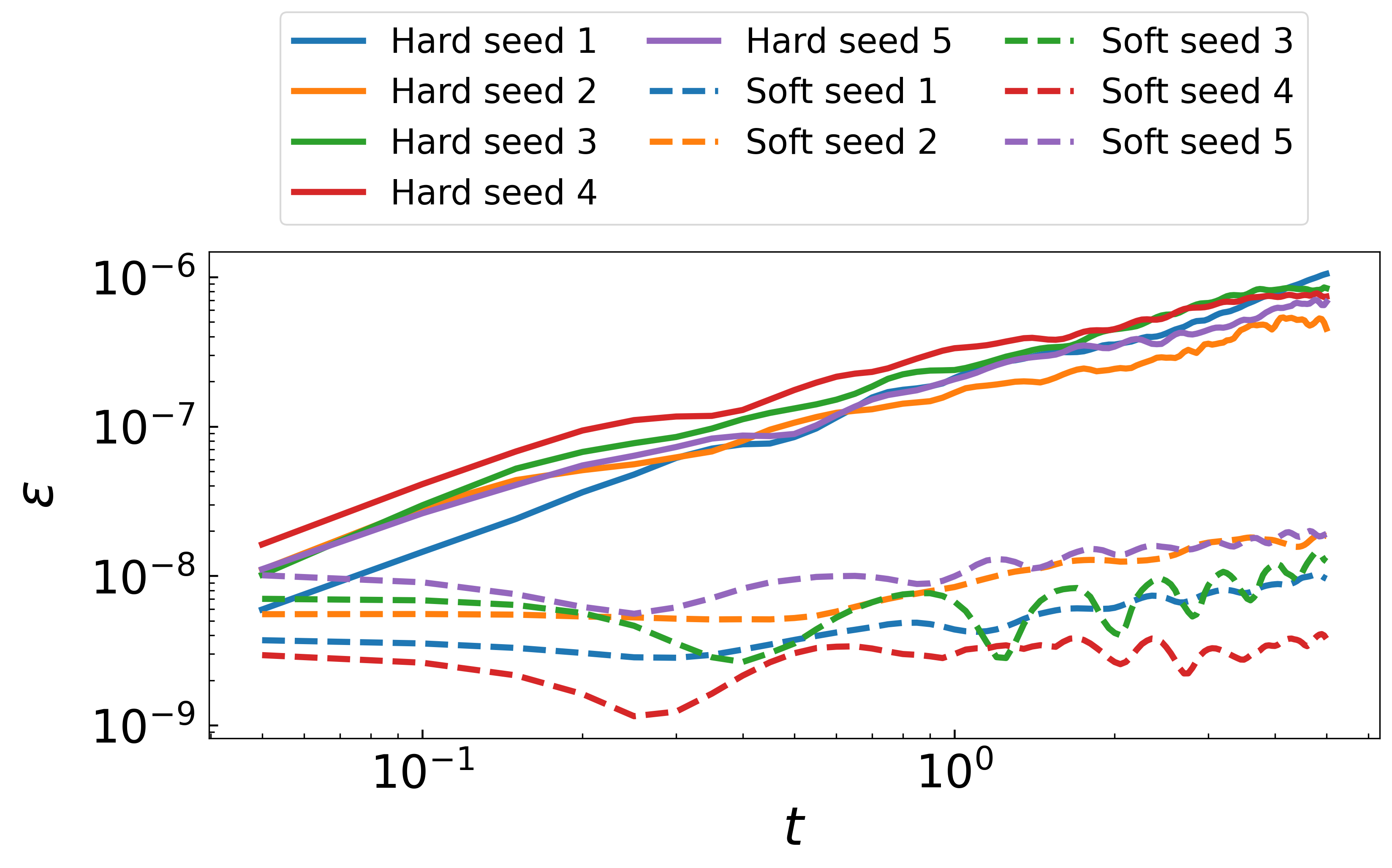}
    \caption{Comparison of hard (solid lines) and soft (dashed lines) enforcement of the temporal initial data for the one-dimensional wave equation. Each color corresponds to an independent random initialization. The curves show the spatially averaged absolute error with respect to the analytical solution. For the hard ansatz used here, the error grows systematically with time for all seeds, whereas soft enforcement maintains a substantially smaller and approximately bounded error throughout the evolution. } 
    \label{fig:hard_soft_comparison}
\end{figure}

Figure ~\ref{fig:hard_soft_comparison} compares the spatially averaged absolute error for five independent random initializations. The hard formulation shows a systematic, approximately monotonic growth of the error in every run. By contrast, the soft formulation maintains an error of order $10^{-9}$--$10^{-8}$ throughout most of the evolution and remains approximately two orders of magnitude more accurate at late times.
For the hard ansatz in Eq.~\eqref{eq:hard_temporal_ansatz}, any approximation error in the neural correction is multiplied by $(t-t_0)^2$. One might therefore suspect that this particular construction is responsible for the loss of long-time accuracy. However, we tested several hard-enforcement constructions, and all of those considered here led to a clear deterioration of the long-time accuracy. We therefore choose to impose the temporal initial data softly in all evolutions presented in this work. This does not compromise the hard enforcement of spatial periodicity through the trigonometric input embedding described in Sec.~\ref{sec:pinns}.

%%%%%%%%%%%%%%%%%%%%%%%%%%%%%%%%%%
\section{Model configurations and computational cost}
\label{app:modelos}
%%%%%%%%%%%%%%%%%%%%%%%%%%%%%%%%%%

Table~\ref{tab:modelos_comparativa} summarizes the main configurations used
for the numerical experiments presented in this work.
For each test, we report the network architecture, temporal decomposition,
number of collocation and initial-data points per time window,
sampling strategy, hardware, and total wall-clock training time.
All temporal initial conditions are imposed softly. Spatial
periodicity in the vacuum benchmarks is enforced exactly through the
trigonometric input embedding described in Sec.~\ref{subsec:conditions}. All models are
trained using the SSBroyden optimizer following the prescription of
Ref.~\cite{urban2025unveiling}, unless stated otherwise.
For all tests, we used a fixed set of collocation points, which was resampled every 100 training epochs.

\begin{table*}[t]
\centering
\begin{tabular}{lcccccc}
\hline
\textbf{Parameter}
& \textbf{L.W.}
& \textbf{G.W.}
& \textbf{G.W. Par.}
& \textbf{Gow. Exp.}
& \textbf{Gow. Col.} 
& \textbf{B.S.} \\
\hline

Input repr.
& Emb.
& Emb.
& Emb.
& Emb.
& Emb.
& $t,x,y,z$ \\

Hidden layers
& 6
& 6
& 6
& 6
& 6
& 6 \\

Neurons/layer
& 22
& 22
& 22
& 22
& 27
& 40 \\

Final time
& 100
& 100
& 100
& 8
& 28
& 400 \\

Time windows
& 50
& 50
& 50
& 4
& 28
& 20 \\

Colloc. pts.
& 1500
& 1500
& 1900
& 1000
& 5000
& 50000 \\

ID pts.
& 1500
& 1500
& 1500
& 1000
& 5000
& 5000 \\
$\lambda_{\text{ID}}$
& 100
& 10
& 10
& 10 
& 1000
& 10 \\

$\lambda_{\text{PDE}}$
& 10
& 1
& 1
& 1
& 10
& 1 \\

$\lambda_{\text{BC}}$
& -
& -
& -
& -
& -
& 1 \\

Sampling
& Random
& Random
& Random
& Random
& Time-exp.
& $\substack{\text{Gaussian (space)}\\\text{Exponential (time)}}$ \\

\hline

GPU
& RTX 4060
& B200
& B200
& RTX 4070 Ti
& B200
& H100 \\

Wall-clock
& 5 h 12 min
& 3 h 24 min
& 8 h 8 min
& 34 min
& 5 h 6 min
& 29 h 28 min \\

\hline
\end{tabular}
\caption{
Main numerical configurations used in the experiments.
The vacuum tests are defined on a periodic domain with the nontrivial
spatial coordinate in $[-1,1]$. The boson-star calculation is
performed on a spherical domain of radius $R_{\rm domain}=60$ using
Cartesian coordinates. The numbers of collocation and initial-data
points refer to each temporal window. All temporal initial conditions
are imposed through soft enforcement. The wall-clock times correspond
to the complete training procedure and are not normalized across GPU
architectures. For comparison, the corresponding finite-difference 
simulation of the Boson star took about 8 hours on 8 CPU cores (Intel Core i7).}
\label{tab:modelos_comparativa}
\end{table*}

%\begin{table*}[t]
%\centering
%\begin{tabular}{lccccccc}
%\hline
%\textbf{Parameter}
%& \textbf{Base}
%& \textbf{$L$}
%& \textbf{$N$}
%& \textbf{$TP_1$}
%& \textbf{$TP_2$} 
%& \textbf{$P_1$} 
%& \textbf{$P_2$} \\
%\hline
%Hidden layers
%& 6 & 3 & 6 & 6 & 6 & 6 & 6 \\
%Neurons/layer
%& 22 & 22 & 11 & 15 & 32 & 22 & 22 \\
%Colloc. pts.
%& 1500 & 1500 & 1500 & 1500 & 1500 & 750 & 3000 \\
%ID pts.
%& 1500 & 1500 & 1500 & 1500 & 1500 & 750 & 3000  \\
%\hline
%Wall-clock
%& 3 h 23 min & 3 h 16 min & 3 h 36 min & 3 h 48 min & 4 h 40 min & 3 h 34 min & 4 h 10 min \\
%\hline
%\end{tabular}
%\caption{
%PINN configurations and wall clock training times for the gauge wave experiments presented in Fig.~\ref{fig:gauge_convergence}. The baseline model is compared against variants featuring half as many hidden layers ($L$), half as many neurons per layer ($N$), half ($TP_1$) or twice ($TP_2$) the number of trainable parameters, and half ($P_1$) or twice ($P_2$) the initial and collocation points.}
%\label{tab:modelos_comparativa_2}
%\end{table*}

Since the experiments were performed on different GPU architectures, the
reported wall-clock times are not intended to provide a direct comparison of
the computational complexity of the different tests. Rather, they provide an
estimate of the computational resources required for the experiments reported
in this work.

%For the collapsing Gowdy test, training was continued through all time windows up to $t=100$, although the PINN ceases to accurately reproduce the expected dynamics at approximately $t\simeq26$, as discussed in the main text. Consequently, the wall-clock time reported in Table~\ref{tab:modelos_comparativa} corresponds to the complete run up to
%$t=100$ %rather than only to the interval over which the solution remains accurate. \FA{xq ponemos este tiempo y no el de t=26}

%\JAP{No entiendo para nada esta tabla. Qué significa que uno reduce a la mitad capas o neuronas, o collocations points a un cuarto,  y el computational time siempre es el mismo ?? Si es porque se ha dejado correr hasta que alcanza la misma precisión (que no lo parece por la comparativa de errores), hay que explicarlo bien, si no, qué es exactamente lo que está pasando ??. Además, lo de llegar a $t=100$ tampoco le veo el sentido. Esta comparación debería hacerse muy diferente, o mejor no hacerla. Tal y como está solo confunde al lector.}

\clearpage
\bibliographystyle{unsrt}
\bibliography{ref}

\end{document}